\documentclass[10pt, conference, letterpaper]{IEEEtran}
\IEEEoverridecommandlockouts

\usepackage{graphicx}
\usepackage{multirow}
\usepackage{subcaption}
\usepackage{tabularx}
\usepackage{booktabs}
\usepackage[nospace]{cite}
\usepackage{amsmath,amssymb,amsfonts}
\usepackage{textcomp}
\usepackage{xcolor}
\usepackage{url}
\usepackage{algorithm}
\usepackage{algorithmicx}
\usepackage{algpseudocode}
\usepackage{amsthm}
\usepackage{hyperref}
\usepackage{booktabs} 
\usepackage{siunitx} 
\usepackage{multirow} 

\begin{document}

\title{HierMoE: Accelerating MoE Training with Hierarchical Token Deduplication and Expert Swap
}

\author{\IEEEauthorblockN{Wenxiang Lin\IEEEauthorrefmark{2}, Xinglin Pan\IEEEauthorrefmark{3}, Lin Zhang\IEEEauthorrefmark{4}, Shaohuai Shi\IEEEauthorrefmark{2}\IEEEauthorrefmark{1}\thanks{*Corresponding author.}, Xuan Wang\IEEEauthorrefmark{2},  Xiaowen Chu\IEEEauthorrefmark{3}\IEEEauthorrefmark{4}\\}
\IEEEauthorblockA{
\IEEEauthorrefmark{2}School of Computer Science and Technology, Harbin Institute of Technology, Shenzhen\\
\IEEEauthorrefmark{3}Data Science and Analytics Thrust, The Hong Kong University of Science and Technology (Guangzhou)\\
\IEEEauthorrefmark{4}Department of Computer Science and Engineering, The Hong Kong University of Science and Technology\\
wenxianglin@stu.hit.edu.cn, xpan413@connect.hkust-gz.edu.cn, lzhangbv@connect.ust.hk  	\\shaohuais@hit.edu.cn, wangxuan@cs.hitsz.edu.cn, xwchu@ust.hk}
}

\maketitle

\newtheorem{theorem}{Theorem}
\newtheorem{corollary}{Corollary}[theorem]
\newtheorem{lemma}{Lemma}[theorem]
\newtheorem{sublemma}{Lemma}[lemma]
\algrenewcommand\algorithmicrequire{\textbf{Input:}}
\algrenewcommand\algorithmicensure{\textbf{Output:}}

\begin{abstract}
The sparsely activated mixture-of-experts (MoE) transformer has become a common architecture for large language models (LLMs) due to its sparsity, which requires fewer computational demands while easily scaling the model size. In MoE models, each MoE layer requires to dynamically choose tokens to activate particular experts for computation while the activated experts may not be located in the same device or GPU as the token. However, this leads to substantial communication and load imbalances across all GPUs, which obstructs the scalability of distributed systems within a GPU cluster. To this end, we introduce HierMoE to accelerate the training of MoE models by two topology-aware techniques: 1) token deduplication to reduce the communication traffic, and 2) expert swap to balance the workloads among all GPUs. To enable the above two proposed approaches to be more general, we build theoretical models aimed at achieving the best token duplication and expert swap strategy under different model configurations and hardware environments. We implement our prototype HierMoE system atop Megatron-LM and conduct experiments on a 32-GPU cluster with DeepSeek-V3 and Qwen3-30B-A3B models. Experimental results show that our HierMoE achieves $1.55\times$ to $3.32\times$ faster communication and delivers $1.18\times$ to $1.27\times$ faster end-to-end training compared to state-of-the-art MoE training systems, Tutel-2DH, SmartMoE, and Megatron-LM. 
\end{abstract}

\begin{IEEEkeywords}
Distributed Deep Learning; Mixture-of-Experts; Expert Parallelism; Token Deduplication; Expert Swap.
\end{IEEEkeywords}

\section{Introduction}

The mixture-of-experts (MoE) architecture with sparse activation has gained significant research interest in large language models (LLMs)~\cite{ DBLP:Shazeer2017outrageously, DBLP:Lepikhin2021gshard,jiang2024mixtral, yang2025qwen3,deepseekv2,jetmoe,liu2024deepseek}. It provides an effective solution for model size scaling, where the computational requirement grows sub-linearly with increasing number of parameters. The MoE architecture incorporates the MoE layer, which comprises multiple feed-forward networks (FFNs), known as \textit{experts}, substituting the dense feed-forward layer while activating only a subset of these experts for each input token~\cite{DBLP:Lepikhin2021gshard,liu2024deepseek}. 
A trainable routing function, generally a small neural network utilizing a softmax mechanism, is employed to dynamically select which experts should be trained for each input token~\cite{DBLP:Lepikhin2021gshard}. This architecture allows the model size to expand to nearly $E$ times (where $E$ represents the number of experts per MoE layer) that of a standard dense model, yet the computational demand remains comparatively stable. 
However, training MoE LLMs typically requires expert parallelism (EP)~\cite{DBLP:Shazeer2017outrageously, hwang2023tutel} to place different experts on different GPUs since a single GPU has limited memory to hold all experts. Due to the dynamic nature of dispatching input tokens to experts that are located in different GPUs, EP introduces significant communications, which are implemented by the AlltoAll collective, easily limiting the scalability of the distributed training system. Recent research~\cite{DBLP:Lepikhin2021gshard, hwang2023tutel, DBLP:liu2022gating, li2023lina, shi2024schemoe, liu2024deepseek,pan2025fsmoe} suggests that communication overheads of the AlltoAll operation constitute 30-60\% of the overall training time in GPU/TPU clusters. Some studies are trying to address the communication problem through 1) algorithmic optimization~\cite{lewis2021base,zhou2022mixture,DBLP:Fedus2022switch,chi2022representation,puigcerver2023sparse} by using better routing functions to balance the workload of experts, and 2) system-level optimization by designing more communication-efficient AlltoAll collective algorithms~\cite{DBLP:Rajbhandari22deepspeedmoe,hwang2023tutel,aminabadi2022deepspeed,DBLP:ma2022bagualu} and adaptive task scheduling to overlap communication tasks and computation tasks~\cite{hwang2023tutel,he2022fastermoe,shi2023pipemoe,li2023lina,zhai2023smartmoe,liu2023janus,pan2024parm, shi2024schemoe, chen2024centauri, jiang2024lancet,pan2025fsmoe,wang2025spmoe,pan2025mitiga,lin2025mast,lin2025scheinfer}. Since the process of algorithmic optimization can significantly impact model convergence, this study concentrates on system-level optimization that does not compromise model accuracy.

\begin{figure}[!t]
	\centering
 \begin{subfigure}[b]{0.24\textwidth}
		\centering
		\includegraphics[width=\textwidth]{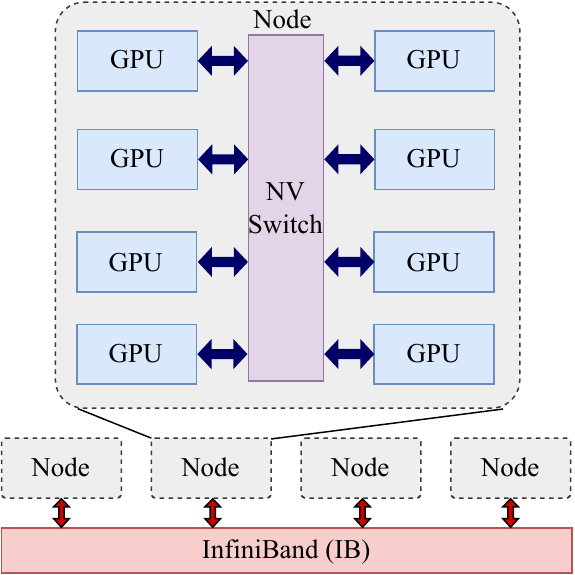}
		\caption{Two-level hierarchy}
		\label{fig:nvswitch}
	\end{subfigure}
    	 \begin{subfigure}[b]{0.24\textwidth}
		\centering
		\includegraphics[width=\textwidth]{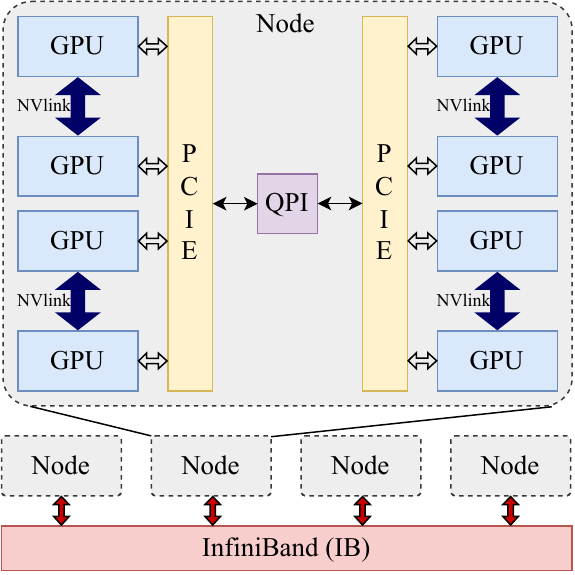}
		\caption{Four-level hierarchy}
		\label{fig:topology}
	\end{subfigure}
	\caption{Two commonly used hierarchical topologies.}
	\label{fig:hiera}
\end{figure}

Specifically, AlltoAll requires each GPU to exchange data with all the other GPUs, so its performance is highly affected by the network topology of GPUs (i.e., the hierarchical connection between GPUs as shown in Fig.~\ref{fig:hiera}). That is, a low-bandwidth link may significantly slow down the overall communication performance. For example, in the four-level hierarchical topology as shown in Fig.~\ref{fig:topology} (Inter-node through InfiniBand, Intra-node through NVLink, PCIe, and QPI), InfiniBand or QPI would easily limit the communication performance of AlltoAll. In larger-scale clusters, GPU nodes should be connected across switches, which introduces higher levels of the topology~\cite{alizadeh2010data,al2008scalable,wang2023topoopt,zhao2025insights}. Existing related optimizations include 1) hierarchical AlltoAll algorithms like two dimensional hierarchical (2DH) AlltoAll in Tutel (Tutel-2DH)~\cite{hwang2023tutel}, PipeA2A in ScheMoE~\cite{shi2024schemoe}, and dedicated kernels for Nvidia Hopper architecture in DeepSeek-V3~\cite{liu2024deepseek} to better utilize Intra-node and Inter-node network bandwidths, and 2) expert placement algorithms to balance the communication workloads of different GPUs like SmartMoE~\cite{zhai2023smartmoe}. \textit{These methods under-estimate the impacts of the hierarchical structure of GPU connection and have not explored the full hierarchical structure to optimize AlltoAll communication, thus achieving suboptimal training performance.}

To this end, we propose HierMoE to fully utilize the hierarchical structure to optimize token distribution and expert migration among GPUs, achieving minimal AlltoAll communication time in MoE model training. HierMoE incorporates three innovative strategies: 
1) conducting theoretical research on the links between hierarchical dimensions and the redundant transfer challenge to design a hierarchical token deduplication AlltoAll algorithm aimed at decreasing data transfer redundancy among varying hierarchical dimensions, 
2) designing a hierarchical expert swap mechanism to balance the communication workloads of different GPUs aimed at further improving the AlltoAll communication efficiency, and 
3) devising theoretical frameworks that render the token deduplication and expert swap strategy broadly applicable and practical for varying models. 
We implement our HierMoE atop the widely-used LLM training system Megatron-LM\footnote{\url{https://github.com/NVIDIA/Megatron-LM/}}, and conduct extensive experiments on a 32-GPU cluster using representative real-world MoE models, including DeepSeek-V3~\cite{liu2024deepseek} and Qwen3-30B-A3B~\cite{yang2025qwen3}. Experimental results show that HierMoE improves the AlltoAll communication efficiency by $1.55\times$ to $3.32\times$ and 
achieves $1.18\times$ to $1.27\times$ faster end-to-end training over the state-of-the-art MoE training systems Tutel-2DH\cite{hwang2023tutel}, SmartMoE\cite{zhai2023smartmoe} and Megatron-LM.
\begin{table}[!h]
	\centering
	\caption{Notations.}
	\label{tab:notations}
\begin{tabular}{ll}
\toprule 
\textbf{Notation} & \textbf{Description} \\
\midrule
$M$ & Embedding dimension of a token \\
$K$ & Number of experts selected for each token \\
$G$ & Number of workers (or GPUs) in the cluster \\
$D$ & Number of hierarchical dimensions \\
$U[i]$ & Number of experts group when \\
&performing Inter-level-$i$ AlltoAll\\
$E$ & Total number of experts \\
$t_{d}$ & Time of $d$-dimensional deduplication hierarchical AlltoAll\\


\bottomrule
\end{tabular}
\end{table}

\section{Preliminaries and Motivations}
In this section, we present some preliminaries of MoE training and our motivations. For convenience, we provide a summary of the commonly used notation in Table~\ref{tab:notations}. 
\begin{figure}[!t]
    \begin{subfigure}[b]{0.48\textwidth}
		\centering
		\includegraphics[width=\textwidth]{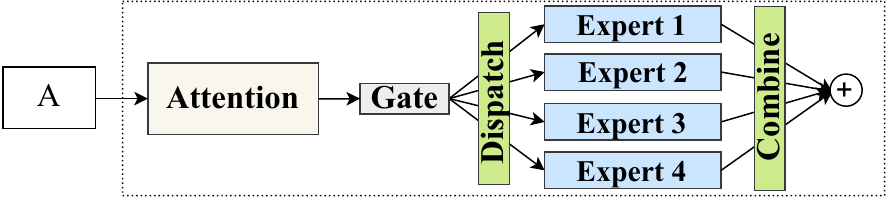}
		\caption{MoE layer}
		\label{fig:moe}
	\end{subfigure}
	 \begin{subfigure}[b]{0.48\textwidth}
		\centering
		\includegraphics[width=\textwidth]{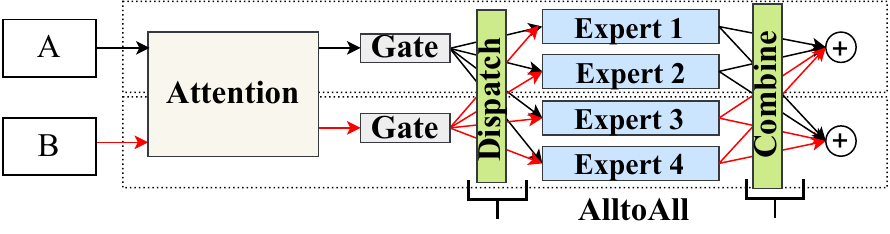}
		\caption{Expert parallelism}
		\label{fig:ep}
	\end{subfigure}
    	\caption{A typical structure of the MoE model.}
	\label{fig:moelayer}
\end{figure}
\subsection{MoE Layer}\label{sec2.1} 
Generally, an MoE model follows the architecture of a dense model by replacing its dense feed-forward layers with MoE layers as shown in Fig.~\ref{fig:moe}.
An MoE layer consists of two primary elements: a gating function and a collection of trainable feed-forward neural networks (FFNs) called experts. At each training iteration, the input tokens are distributed to selected experts according to the gating function. The gating function also uses a small trainable FFN followed by a softmax layer and top-$K$ selection to determine which experts should process which tokens. 

\subsection{Expert Parallelism}
Since a single GPU has limited memory to accommodate all experts, the expert parallelism (EP)~\cite{DBLP:Shazeer2017outrageously,DBLP:Lepikhin2021gshard,hwang2023tutel} is required to place different experts at each layer to different GPUs during training. Let $E$ indicate the number of experts at each MoE layer and $G$ the number of GPUs in the cluster. EP would place $E/G$ experts every MoE layer on each GPU.
Together with data parallelism (DP)~\cite{DBLP:dean2012large,Megatronlm}, which is a de facto training paradigm in distributed training of LLMs, the input data in each device is different. Therefore, the token distribution generated by the gating function requires an AlltoAll collective method to dispatch tokens to particular experts (referred to as \textit{AlltoAll Dispatch}), illustrated as ``Dispatch'' shown in Fig.~\ref{fig:ep}. It ensures that data is routed to the correct experts for processing. After dispatch, the tokens undergo processing by their designated experts. The results from the experts then need to be processed by subsequent layers (e.g., Attention) of the MoE layer, requiring another AlltoAll operation to merge the expert outputs (termed as \textit{AlltoAll Combine}) illustrated as ``Combine'' shown in Fig.~\ref{fig:ep}. The two AlltoAll operations at every MoE layer generally introduce significant communication overheads, which limit the scaling efficiency of training systems. Modern systems like Tutel~\cite{hwang2023tutel} and DeepSeed-MoE~\cite{DBLP:Rajbhandari22deepspeedmoe} try to use the two-dimensional hierarchical AlltoAll (2DH-AlltoAll) algorithm~\cite{DBLP:Rajbhandari22deepspeedmoe,hwang2023tutel} that is a dedicated design for the hierarchical topology of GPUs. 
\subsection{Motivations}

\begin{figure}[!t]
	\centering
 \begin{subfigure}[b]{0.24\textwidth}
		\centering
		\includegraphics[width=\textwidth]{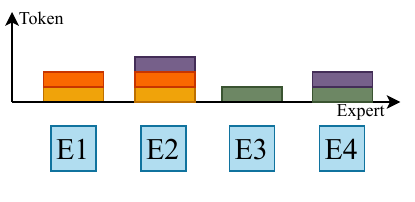}
		\caption{Tokens assigned to each expert}
		\label{fig:pe}
	\end{subfigure}
	 \begin{subfigure}[b]{0.24\textwidth}
		\centering
		\includegraphics[width=\textwidth]{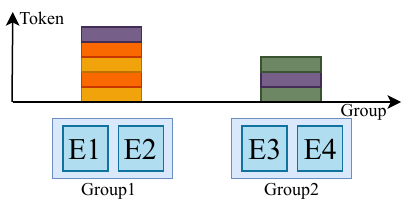}
		\caption{Tokens assigned to each group}
		\label{fig:pg2}
	\end{subfigure}
     \begin{subfigure}[b]{0.24\textwidth}
		\centering
		\includegraphics[width=\textwidth]{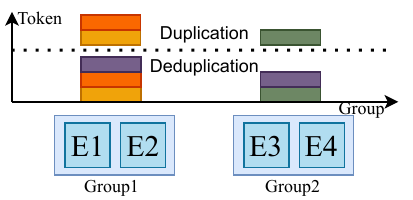}
		\caption{Tokens assigned to each group after deduplication}
		\label{fig:pd}
	\end{subfigure}
  \begin{subfigure}[b]{0.24\textwidth}
		\centering
		\includegraphics[width=\textwidth]{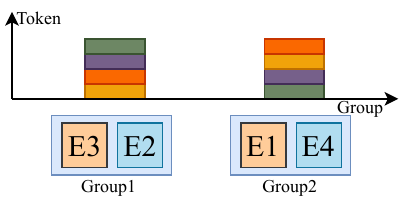}
		\caption{Tokens assigned to each group after expert swap}
		\label{fig:ps}
	\end{subfigure}
	\caption{The illustration of tokens assignment with different configurations. Different colors represent different tokens.}
	\label{fig:psample}
\end{figure}

\begin{table}[]
	\centering

 		\caption{Token duplication rates with different $K$ and $R$.}
		\label{tab:motivation}

\begin{tabular}{lcccc}
\toprule
 & \multicolumn{4}{c}{\textbf{$K$}} \\
 \cmidrule(lr){2-5}
\textbf{$R$}  & 2 & 4 & 6 & 8 \\
\midrule
32 & 2\% & 4\% & 7\% & 9\% \\
16 & 3\% & 9\% & 14\% & 18\% \\
8 & 6\% & 17\% & 27\% & 34\% \\
4 & 12\% & 32\% & 46\% & 55\% \\
\bottomrule
\end{tabular}
\end{table}

\subsubsection{Duplicate tokens with the hierarchical topology}

Since each GPU would hold $E/G$ experts per MoE layer when the number of experts is larger than the number of GPUs (i.e., $E/G>1$), tokens are required to be \textit{redundantly} transmitted to multiple selected experts that are located on the same GPU to exploit the AlltoAll collective.
As illustrated in Fig.~\ref{fig:pe}, each expert is assigned particular tokens, and a single token requires multiple experts (i.e., $K$ in the top-$K$ selection). Assume that each GPU (or group) holds two experts as shown in Fig~\ref{fig:pg2}. Every group would have duplicate tokens, which results in redundant communication in the AlltoAll operation. \textit{Thus, eliminating this duplication can reduce the communication traffic as shown in Fig.~\ref{fig:pd}}.

Moreover, the duplication rate is highly affected by the number of groups (say $R$) and the number of selected experts (i.e., $K$) per token. We conduct preliminary experiments with different $R$ and $K$ to measure the duplication rate at each group as shown in Table~\ref{tab:motivation}. The results indicate that lower $R$ (The hierarchical topology can divide experts into different groups.) and higher $K$ (which is very common in modern MoE models like DeepSeek-V3, Qwen-MoE, etc.) would result in a higher duplication rate. 
\textit{Thus, how to eliminate the duplicated tokens by considering $K$, $R$, and the GPU topology becomes more challenging.}

\subsubsection{Unbalanced routing workloads with the hierarchical topology}
Since the selected experts for each token are determined by the routing function, it is easy to cause imbalanced workloads for each expert, which results in increased communication traffic~\cite{DBLP:Lepikhin2021gshard,liu2024deepseek,zhai2023smartmoe}. 
Existing solutions like FlexMoE~\cite{nie2023flexmoe} and SmartMoE~\cite{zhai2023smartmoe} dynamically adjust expert placement during training to balance token distribution across GPUs, \textit{but they neither account for token deduplication nor adapt to the hierarchical topology of GPUs.} If the token duplications have been overlooked, simply swapping experts to balance the workload could result in a higher communication overhead. For example, as shown in Fig.~\ref{fig:ps}, we swap expert 1 and expert 3 such that the workload of each group is more balanced, but its communication traffic becomes higher than that of Fig.~\ref{fig:pd}. 


\textit{Therefore, it requires a new expert swap strategy taking into account token deduplication and hierarchical bandwidth constraints to achieve higher training performance.}
\section{Hierarchical Token Deduplication}
\begin{figure}[!t]
	\centering
 \begin{subfigure}[b]{0.48\textwidth}
		\centering
		\includegraphics[width=\textwidth]{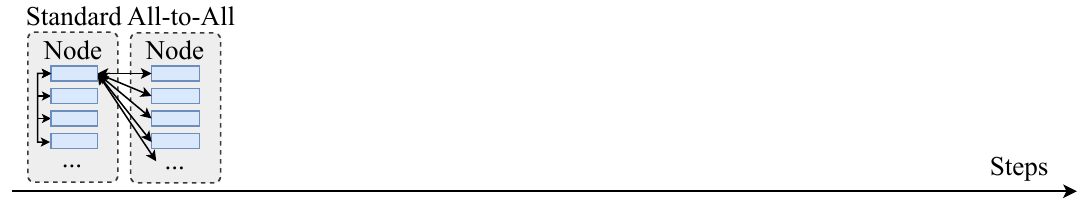}
		\caption{One dimension}
		\label{fig:LEVEL0}
	\end{subfigure}
    
	 \begin{subfigure}[b]{0.48\textwidth}
		\centering
		\includegraphics[width=\textwidth]{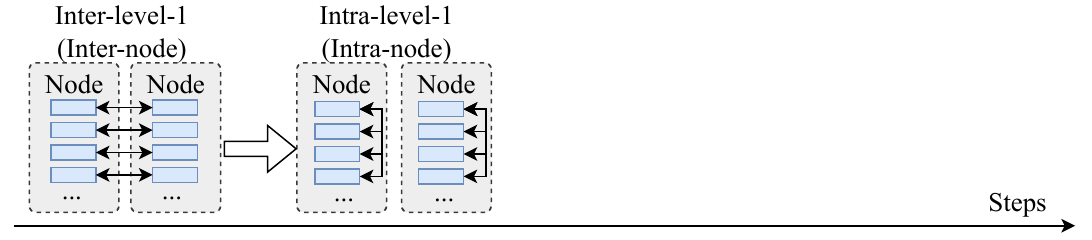}
		\caption{Two dimensions}
		\label{fig:LEVEL1}
	\end{subfigure}
    
  \begin{subfigure}[b]{0.48\textwidth}
		\centering
		\includegraphics[width=\textwidth]{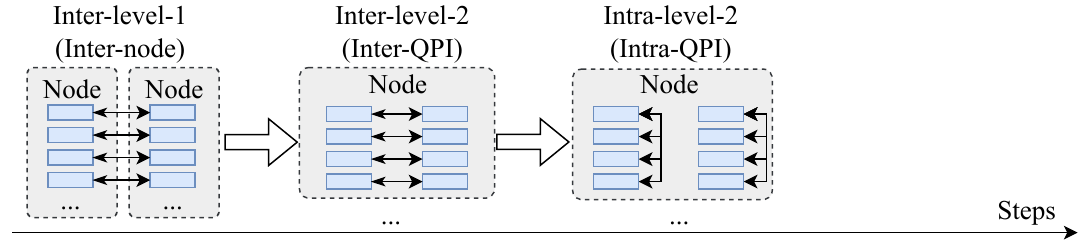}
		\caption{Three dimensions}
		\label{fig:LEVEL2}
	\end{subfigure}
    
   \begin{subfigure}[b]{0.48\textwidth}
		\centering
		\includegraphics[width=\textwidth]{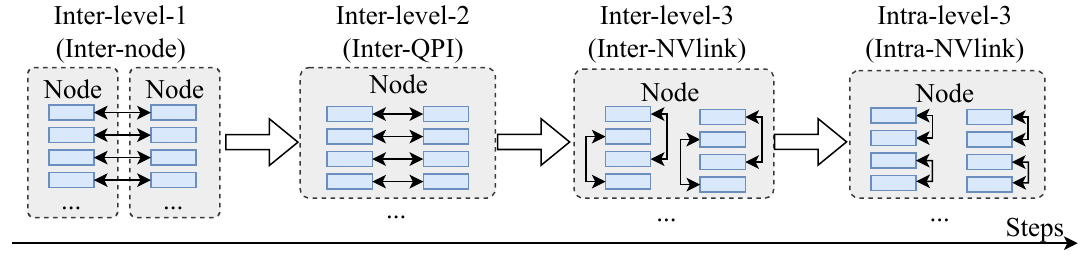}
		\caption{Four dimensions}
		\label{fig:LEVEL3}
	\end{subfigure}
	\caption{Four types of hierarchical AlltoAll with different dimensions. The example has two nodes with eight GPUs per node. We use ``...'' to omit some GPUs and nodes.}
	\label{fig:LEVELS}
\end{figure}

\subsection{Hierarchical Deduplication AlltoAll}  


To better utilize the hierarchical topology for token transferring in the MoE layer, we design a multi-dimensional AlltoAll algorithm with token deduplication, called HierD-AlltoAll. To make our design general to existing AlltoAll algorithms, existing standard AlltoAll and 2DH-AlltoAll algorithms can be seen as particular cases. Specifically, for the standard AlltoAll algorithm, it can be denoted as a one-dimensional algorithm as it does not consider any topology, as shown in Fig.~\ref{fig:LEVEL0}. Similarly, the 2DH-AlltoAll algorithm is a two-dimensional algorithm that is dedicated for two-dimensional hierarchical topology as shown in Fig.~\ref{fig:LEVEL1}.
For intricate topologies with more than two layers of hierarchy, we arrange the GPUs into groups, ensuring that the number of groups aligns with the hierarchy levels. For example, for the four levels of hierarchy, like the common case where each node has NVLink, PCIe, and QPI connections as shown in Fig.~\ref{fig:topology}, we organize GPUs into four groups and do a four-dimensional AlltoAll. 
As shown in Fig.~\ref{fig:LEVEL3}, the first level (Inter-level-1) simultaneously invokes 8 Inter-node AlltoAll operations, each pair of GPUs communicates with each other through IB. The second level (Inter-level-2) simultaneously invokes 8 Inter-QPI AlltoAll, each of which only has two GPUs. Similarly, the third level (Inter-level-3) performs Inter-NVLink AlltoAll, and the fourth level (Intra-level-3) invokes Intra-NVLink AlltoAll to complete the functionality of the original AlltoAll. In general, a ($d$)-dimensional hierarchical AlltoAll is composed of Inter-level-1, Inter-level-2, up to Inter-level-($d$-1) AlltoAll followed by an Intra-level-($d$-1) AlltoAll. As for an ($d$+1)-dimensional AlltoAll algorithm, we further split the Intra-level-($d$-1) AlltoAll into Inter-level-($d$) and Intra-level-($d$) AlltoAll.

\begin{figure}[!t]
	\centering
		\includegraphics[width=0.48\textwidth]{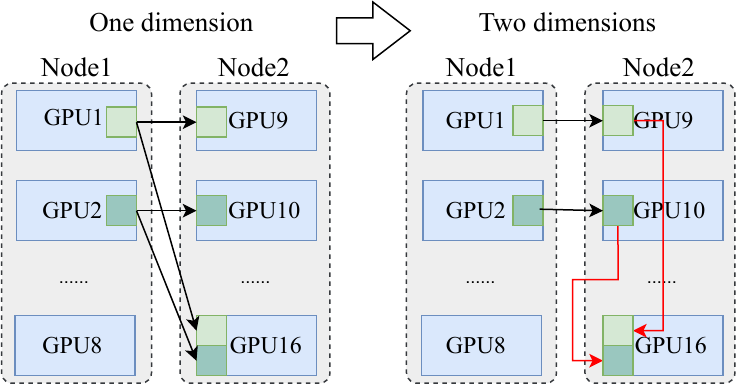}
	\caption{An illustration of token deduplication in hierarchical AlltoAll with 2 nodes and 16 GPUs. }
	\label{fig:token-dedup}
\end{figure}

\begin{figure}[!t]
	\centering
		\includegraphics[width=0.48\textwidth]{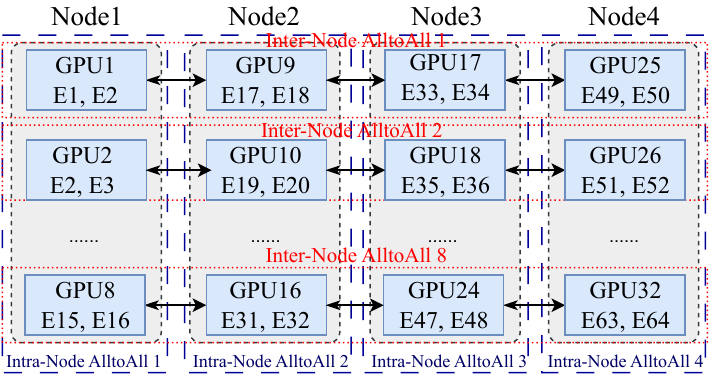}
	\caption{An illustration of experts and GPUs index for Inter-Node/Intra-Node AlltoAll. }
	\label{fig:index}
\end{figure}

According to the hierarchical AlltoAll algorithm, we design token deduplication strategy for minimizing the overall time of communication and propose Hier-AlltoAll. Let $D$ denote the number of dimensions for the hierarchical structure. As the high-dimensional hierarchical topology can also perform low-dimensional hierarchical AlltoAll, we refer to the deduplication version of $D$ kinds of dimensional hierarchical AlltoAll shown in Fig.~\ref{fig:LEVELS} as HD1-AlltoAll, HD2-AlltoAll, and up to HD$D$-AlltoAll, respectively.

We take HD1-AlltoAll and HD2-AlltoAll as an example to demonstrate the initiative effect of the hierarchical deduplication.
 As illustrated in Fig.~\ref{fig:token-dedup}, the HD2-AlltoAll (shown on the right) with deduplicated tokens in both dimensions requires only two tokens transferred to another node using Inter-node communication. In contrast, the HD1-AlltoAll (depicted on the left) necessitates the dispatch of four tokens. The Inter-node AlltoAll from HD2-AlltoAll redistributes experts from 16 groups by GPUs into 2 groups by nodes, leading to increased duplicated tokens in each group. By removing these duplications, we improve Inter-node communication traffic over IB, which has low bandwidth. But HD2-AlltoAll requires two more tokens transferred to other GPUs using Intra-node communication. Similarly, using HD3-AlltoAll can lower the communication traffic of Inter-QPI communication through QPI compared to HD2-AlltoAll, but increase the communication volume of Intra-QPI communication. Notably, the experts group number of Inter-QPI AlltoAll is bigger than that of Inter-Node AlltoAll, as it further splits the experts group by QPI. According to the Table.~\ref{tab:motivation}, high top-$K$ selection and low experts group results in high duplication rates and vice versa. Token deduplication benefits might diminish when employing HD3-AlltoAll rather than HD2-AlltoAll. The same applies to HD4-AlltoAll. 
 
 Therefore, the dimension of the hierarchical AlltoAll is not necessarily larger. We need to formulate the performance model of different dimensional AlltoAll to determine the optimal dimension, ensuring communicational overhead reduction. 



\subsection{Performance Model}
We model the time cost of the standard AlltoAll communication (also HD1-AlltoAll) via linear models~\cite{shi2023pipemoe} 
as follows (will verify in \S\ref{subsec:perfmodel-evaluation}):
\begin{equation}\label{pm}
{t_{1}} = {\alpha _{a2a}} + n_{a2a} \cdot {\beta _{a2a}},
\end{equation}
where $n_{a2a}$ represents the volume of the communication message, ${\alpha _{a2a}}$ denotes the startup time and ${\beta _{a2a}}$ represents the time per byte transmitted. $\alpha$ and $\beta$ parameters of Inter/Intra-level-($i$) AlltoAll are represented as $\alpha_{a2a}^{\text{Inter/Intra}(i)}$ and $\beta_{a2a}^{\text{Inter/Intra}(i)}$ respectively.   

Unlike ${\beta _{a2a}}$ and $\alpha_{a2a}$ associated with the cluster and determined during initialization, $n_{a2a}$ is related to the dynamic routing results of the MoE layer. We further model $n_{a2a}$ as the product of the number of GPUs in the AlltoAll operation and the number of tokens sent to each GPU.
\begin{equation}
    n_{a2a} = G \cdot \max (p) \cdot M \cdot v,
    \label{equ:n0}
\end{equation}
where $G$ denotes the number of GPUs in the cluster, $p \in \mathbb{R}^{G}$ represents the duplicate-free number of tokens assigned to each expert group (the number of groups is the same as that of GPUs in HD1-AlltoAll), $M$ denotes the embedding dimension size of each token and $v$ denotes the bytes of one embedding dimension. To ensure that all tokens are dispatched, we use $\max(p)$ to represent the number of tokens sent to each GPU.

For HD$d$-AlltoAll where $d>1$, as shown in Fig.~\ref{fig:LEVELS}, a ($d$)-dimensional hierarchical AlltoAll is composed of Inter-level-1, Inter-level-2, up to Inter-level-($d$-1) AlltoAll followed by an Intra-level-($d$-1) AlltoAll. We thus formulate the time cost of ($d$)-dimensional AlltoAll as 
\begin{equation}
\begin{aligned}
   t_{d} = &  \sum\limits_{i = 1}^{d-1} (n_{a2a}^{\text{Inter(}i\text{)}}\cdot \beta _{a2a}^{{\text{Inter(}\textit{i}\text{)}}} + \alpha _{a2a}^{{\text{Inter(}\textit{i}\text{)}}}) \\
   & + n_{a2a}^{\text{Intra(}d\text{-1)}} \cdot \beta_{a2a}^{\text{Intra(}d\text{-1)}} + \alpha_{a2a}^{\text{Intra(}d\text{-1)}},\\
\end{aligned}
\label{equ:level}
\end{equation}
where $1 < d \leq D.$ Similar to Eq.~(\ref{equ:n0}), we also use the product of the number of GPUs in an Inter-level-($i$) AlltoAll and the number of tokens sent to each GPU to represent $n^{\text{Inter(}i\text{)}}_{a2a}$. Notably, input tokens of Inter-level-($i$) AlltoAll and the group number of experts $U[i]$ are different from each other. So we distinguish the number of tokens assigned to each expert group for Inter-level-($i$) AlltoAll as $p^{\text{Inter(}i\text{)}}_{a2a}\in \mathbb{R}^{U[i]}$. 
Similarly, we use $\max(p^{\text{Inter(}i\text{)}}_{a2a})$ to represent the number of tokens sent to each GPU during Inter-level-($i$) AlltoAll.

Notably, $U[i]$ in $U \in \mathbb{R}^{D}$ denotes the group number of experts when performing Inter-level-($i$) AlltoAll. Taking the topology shown in Fig.~\ref{fig:topology} as the example, Inter-level-$1$ (Inter-Node) AlltoAll divides experts into four groups by nodes so $U[1]=4$ (also illustrated in Fig.~\ref{fig:index}). Inter-level-$2$ (Inter-QPI) AlltoAll further splits experts in each node into two parts by QPI so $U[2] =8$. Inter-level-$3$ (Inter-NVLink) AlltoAll divides experts in each QPI group into two parts so $U[3] =16$. Specially, we set $U[0]=1$. 

Additionally, through our numerical analysis, we find that $\frac{U[i]}{U[{i-1}]}$ can signify the number of GPUs involved in an Inter-level-($i$) AlltoAll while $\frac{G}{U[{d-1}]}$ can represent the GPUs count participating in an Intra-level-($d$-1) AlltoAll. Interestingly, the number of GPUs used in an Inter/Intra-level AlltoAll differs from that of expert groups. Because both Inter-level-($i$+1) and Intra-level-($i$+1) AlltoAll take place within the GPUs group of an Intra-level-($i$) AlltoAll as they are derived from it. For instance, Inter/Intra-QPI AlltoAll occurs within a node without interfacing with GPUs from other nodes. Consequently, Inter-level-($i$) AlltoAll will first divide experts into $U[i]$ groups to count duplicate-free tokens and then select corresponding $\frac{U[i]}{U[{i-1}]}$ groups to dispatch tokens. 
Then we can derive that
\begin{equation}
    n_{a2a}^{\text{Inter(}i\text{)}} =\frac{U[i]}{U[{i-1}]}\cdot \max(p_{a2a}^{\text{Inter(}i\text{)}}) \cdot M \cdot v.
     \label{equ:ninter}
\end{equation}

Similarly, we use $\max (p_{a2a}^{\text{Intra(}d\text{-1)}})$ to represent the number of tokens sent to each GPU during Intra-level-($d$-1) AlltoAll. Specially, the expert group count is the same as the number of GPUs for all Intra-level AlltoAll. Intra-level-($d$-1) AlltoAll will first divide experts into $G$ groups and then select corresponding $\frac{G}{U[{d-1}]}$ groups to dispatch, given that the index of experts in GPUs selected by Intra-level AlltoAll is always contiguous as shown in Fig.~\ref{fig:index}. So we have
\begin{equation}
     n_{a2a}^{\text{Intra(}d\text{-1)}} =\frac{G}{U[{d-1}]}\cdot \max(p_{a2a}^{\text{Intra(}d\text{-1)}}) \cdot M \cdot v.
     \label{equ:nintra}
\end{equation}

\subsection{Problem Formulation and Solution}
Based on the above performance models for different dimensional hierarchical deduplication AlltoAll, we can derive the problem of determining the optimal dimension $d^*$ as 
\begin{equation}
\begin{aligned}
    {d^*} =\left\{ \begin{array}{c} 1 , \quad{t_{1}} < \mathop {\min }\limits_{1 < d \leq {D}} ({t_{{d}}})\\
\operatorname*{arg\,min}\limits_{1 < d \leq {D}}(t_{d}),\quad else\\
\end{array} \right.
\end{aligned}
\label{equ:optimal}
\end{equation}

All parameters in Eq.~(\ref{equ:n0}), Eq.~(\ref{equ:ninter}), and Eq.~(\ref{equ:nintra}) are cluster-related and can be pre-initialized, except $p$, $p_{a2a}^{\text{Inter(}i\text{)}}$, and $p_{a2a}^{\text{Intra(}d\text{-1)}}$, which need to be calculate by the MoE layer's routing results. To formulate the relationship, we use $p_{a2a}^{(l,g)}  \in \mathbb{R}^{g}$, which denotes the duplicate-free number of tokens assigned to $g$ expert groups of Inter-level-($l$) or Intra-level-($l$-1) AlltoAll, to generally represent $p$ (i.e., $p_{a2a}^{(1,G)}$), $p_{a2a}^{\text{Inter(}i\text{)}}$ (i.e., $p_{a2a}^{(i,U[i])}$) and $p_{a2a}^{\text{Intra(}d\text{-1)}}$ (i.e., $p_{a2a}^{(d,G)}$). Notably, as shown in Fig.~\ref{fig:LEVELS}, input tokens of Inter-level-($d$) AlltoAll are the same as that of Intra-level-($d$-1) AlltoAll so we can use $p_{a2a}^{(d,G)}$ to represent $p^{\text{Intra(}d\text{-1)}}_{a2a}$. Let $\mathcal{I}_{route}^{(l,E)} \in \mathbb{R}^{T'[l]\times E}$ represent the routing result mask for input tokens of Inter-level-($l$) AlltoAll with the datatype of boolean, $T'[l]$ being the number of input tokens of Inter-level-($l$) AlltoAll. And $\mathcal{I}^{(l,E)}_{route}[i,j]$ represents wether the $i$-th token select $j$-th expert. Then we can formulate $p^{(l,g)}_{a2a}[j]$ by 
\begin{equation}
\begin{aligned}
            \mathcal{I}^{(l,g)}_{route}[i,j]&= \bigvee_{j_1 = (j-1)\frac{E}{g} + 1}^{j\cdot \frac{E}{g}} \mathcal{I}^{(l,E)}_{route}[i,j_1], \\
        p_{a2a}^{{(l,{g})}}[j]&= \sum_{i} \mathbb{I}(\mathcal{I}^{(l,g)}_{route})[i,j],
\end{aligned}
\label{routing}
\end{equation}
where $\bigvee$ denotes the bitwise OR operation, allowing for the elimination of deduplication tokens, and $\mathcal{I}^{(l,g)}_{route}[i,j]$ represents whether the $i$-th token selects the $j$-th expert group. 
Denote $T$ as the total number of tokens for the MoE layer, the time complexity to calculate $p$, $p_{a2a}^{\text{Inter(}i\text{)}}$ and $p_{a2a}^{\text{Intra(}d\text{-1)}}$ is $O(D\cdot T\cdot K)$.
Then, we proceed by examining each possible value of $d$ to determine the optimal dimension. HierD-AlltoAll refers to hierarchical deduplication AlltoAll with this optimal $d^*$.

\begin{algorithm}[!t]
\caption{Find the Optimal Dimension for HierD-AlltoAll}
\label{alg:DEDUP_algorithm}
\begin{algorithmic}[1]
\Require{%
    $\mathcal{I}_{route}^{(1,E)},U,M,G,E,D,\beta_{a2a},\alpha_{a2a}$, \\
    $\beta_{a2a}^{\text{Inter(}l\text{)}}, \alpha_{a2a}^{\text{Inter(}l\text{)}},$ 
    $\beta_{a2a}^{\text{Intra(}l\text{)}}, \alpha_{a2a}^{\text{Intra(}l\text{)}}, 0< l < D$%
}
\Ensure{Optimal dimension $d^*$} 
\State $m\gets E/G$
\State $\mathcal{I}_{route}^{(1,G)}[{i,j}] \gets \bigvee_{j_1 = (j-1)m + 1}^{j\cdot m} \mathcal{I}^{(1,E)}_{route}[{i,j_1}], \quad 1\le j \le G$
\State $p[{j}]\gets \sum_{i} \mathbb{I}(\mathcal{I}^{(1,G)}_{route}[{i,j}])$
\For{$0<k<D$}
    \State $m\gets E/U[k]$
    \State $\mathcal{I}^{(k,U[k])}_{route}[i,j] \gets \bigvee_{j_1 = (j-1)m + 1}^{j\cdot m} \mathcal{I}^{(k,E)}_{route}{[i,j_1]}, 1\le j \le U[k]$
    \State $p_{a2a}^{(k,U[k])}[j] \gets \sum_{i} \mathbb{I}(\mathcal{I}^{(k,U[k])}_{route}{[i,j]})$
    \State $\mathcal{I}_{route}^{(k+1,E)} \gets process(\mathcal{I}_{route}^{(k,E)})$\Comment{monitoring the change after performing Inter-level-($k$) communication}
    \State $p_{a2a}^{(k+1,G)}[j] \gets \sum_{i} \mathbb{I}(\mathcal{I}^{(k+1,E)}_{a2a}[{i,j}])$    
\EndFor
\State $d^* \gets \text{Eq. (\ref{equ:optimal})}$ 
\State \textbf{return} $d^*$ 
\end{algorithmic}
\end{algorithm}

\subsection{Algorithm}
According to the above solution, we derive the algorithm to determine the optimal dimension of HierD-AlltoAll for any given MoE layer as shown in Algorithm~\ref{alg:DEDUP_algorithm}. The input including the embedding size of the token $M$, the routing result mask $\mathcal{I}^{(1,E)}_{route}$, the number of GPUs $G$, the number of experts $E$, the number of dimensions for the hierarchical structure in the cluster $D$, the expert number of group $U$ for each Inter-level AlltoAll and cluster parameters   $\beta_{a2a},\alpha_{a2a},$$\beta_{a2a}^{\text{Inter(}l\text{)}}, \alpha_{a2a}^{\text{Inter(}l\text{)}},$  $\beta_{a2a}^{\text{Intra(}l\text{)}},\alpha_{a2a}^{\text{Intra(}l\text{)}}, 0< l < D$. In the algorithm, we first calculate ${p}$ with the mask routing results $\mathcal{I}_{route}^{(1,E)}$ from the MoE layer (Line 2-4). Then, we monitor tokens' changing in $\mathcal{I}_{route}^{(1,E)}$ to get $\mathcal{I}_{route}^{(k,E)}$ of Inter-level-($k$) communication (Line 9)
and calculate $p_{a2a}^{(k+1,G)}$ and $p_{a2a}^{(k,U[k])}$ (Line 6-8 and Line 10). Finally, we get the optimal $d^*$ following Eq. (\ref{equ:optimal}) (Line 12).

\section{Hierarchical Expert Swap}
Our HierD-AlltoAll addresses the token duplication problem, but the workloads of different GPUs may still be imbalanced. 
Earlier methods, such as SmartMoE~\cite{zhai2023smartmoe}, swap experts by counting allocated tokens without considering duplicated tokens to determine the distribution across GPUs, which is incompatible with our proposed HierD-AlltoAll. To address this, we introduce a hierarchical expert swap strategy (HierD-ES) tailored for our HierD-AlltoAll communication that counts duplicate-free tokens assigned to each hierarchical group. Specifically, in HierD-ES, the key idea is to swap the positions of two experts during training, and the two experts are iteratively chosen to minimize communication overhead with the time model of Eq.~(\ref{equ:level}). The main challenge is to formulate the optimization problem and develop the optimal solution with minimal overhead.

\begin{figure}[!t]
	\centering
		\includegraphics[width=0.48\textwidth]{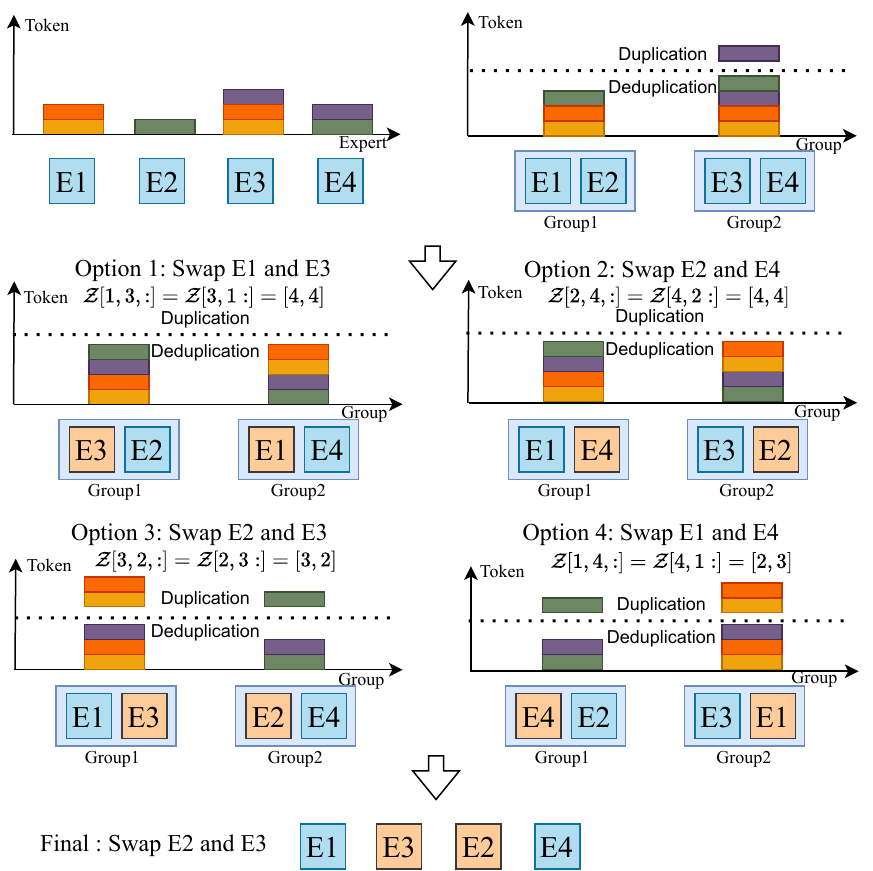}
	\caption{An illustration of our strategies to swap experts. We count the duplicate-free tokens assigned to each group after swapping any two experts and select the expert pair that minimizes the communication overhead.}
	\label{fig:hda}
\end{figure}

\subsection{Problem Formulation and Solution}
HierD-ES needs to count the duplicate-free tokens assigned to each hierarchical group after swapping two experts. So we combine $t_1$ and $t_{d}$ on Eq.~(\ref{pm}) and Eq.~(\ref{equ:level}) and extend to $\mathcal{Q}_d\in \mathbb{R}^{E\times E}$. Each element $\mathcal{Q}_d[r,c]$ represents the estimated time cost of $d$-dimensional hierarchical deduplication AlltoAll after swapping the positions of $r$-th and $c$-th experts. And we formulate $\mathcal{Q}_d$ by $\mathcal{N}^{\text{Inter-}i}_{a2a}\in \mathbb{R}^{E\times E}$ and $\mathcal{N}^{\text{Intra-}\text{(}d\text{-1})}_{a2a}\in \mathbb{R}^{E\times E}$. Each element $\mathcal{N}^{\text{Inter-}i}_{a2a}[r,c]$ and $\mathcal{N}^{\text{Intra-}\text{(}d\text{-1})}_{a2a}[r,c]$ are communication bytes for Inter-level-($i$) and Intra-level-($d$-1) AlltoAll after swapping the positions of $r$-th and $c$-th experts.
\begin{equation}
\begin{aligned}
    \mathcal{Q}_d[r,c] &= \sum\limits_{i = 1}^{d-1} \left( \mathcal{N}_{a2a}^{\text{Inter(}i\text{)}}[r,c] \cdot \beta_{a2a}^{\text{Inter(}\textit{i}\text{)}} + \alpha_{a2a}^{\text{Inter(}\textit{i}\text{)}} \right) \\
    &\quad + \mathcal{N}_{a2a}^{\text{Intra}(d\text{-}1)}[r,c] \cdot \beta_{a2a}^{\text{Intra}(d\text{-}1)} + \alpha_{a2a}^{\text{Intra}(d\text{-}1)}, \\
    & \hspace{13em}   0 < d \leq D  
\end{aligned}
\label{eq:estimate}  
\end{equation}
Specially, we set $\alpha_{a2a}^{\text{Intra(0)}} = \alpha_{a2a}$ and $\beta_{a2a}^{\text{Intra(0)}} = \beta_{a2a}$ to cover one dimensional AlltoAll. And similar to Eq.~(\ref{equ:n0}), Eq.~(\ref{equ:ninter}) and Eq.~(\ref{equ:nintra}), we can formulate $\mathcal{N}_{a2a}^{\text{Inter-}i}[r,c] $ and $\mathcal{N}_{a2a}^{\text{Intra-}(d\text{-}1)}[r,c]$ as the product among the number of GPUs in the corresponding AlltoAll, the number of tokens sent to each GPU, the embedding dimension of a token $M$ and bytes per dimension $v$.
\begin{equation}
\left\{
\begin{aligned}
       \mathcal{N}_{a2a}^{\text{Inter(}i\text{)}}[r,c] &=\frac{U[i]}{U[{i-1}]}\cdot \max_{}(\mathcal{Z}_{a2a}^{\text{Inter(}i\text{)}}[r,c,:]) \cdot M \cdot v,  \\ 
          \mathcal{N}_{a2a}^{\text{Intra}(d\text{-}1)}[r,c]& =\frac{G}{U[{d-1}]}\cdot \max(\mathcal{Z}_{a2a}^{\text{Intra}(d\text{-}1)}[r,c,:]) \cdot M \cdot v.\\
\end{aligned}
\right.
\label{equ:N}
\end{equation}
$\frac{U[i]}{U[{i-1}]}$ and $\frac{G}{U[{d-1}]}$ are the number of GPUs involved in an Inter-level-($i$) AlltoAll and Intra-level-($d$-1) AlltoAll which have been discussed on Eq.~(\ref{equ:ninter}) and Eq.~(\ref{equ:nintra}). We use $ \max_{}(\mathcal{Z}_{a2a}^{\text{Inter-}i}[r,c,:])$ and $\max(\mathcal{Z}_{a2a}^{\text{Intra-}(d\text{-}1)}[r,c,:])$ to represent the number of tokens sent to each GPU in Inter-level-($i$) AlltoAll and Intra-level-($d$-1) AlltoAll after swapping $r$-th and $c$-th experts. And each element $\mathcal{Z}_{a2a}^{\text{Inter-}i}[r,c,k]$ of $\mathcal{Z}^{\text{Inter-}i}_{a2a}\in \mathbb{R}^{E\times E\times U[i]}$ denotes the duplicate-free number of tokens assigned to $k$-th expert group of size $U[i]$ after swapping the positions of $r$-th and $c$-th experts given that Inter-level-($i$) AlltoAll will first categorize experts into $U[i]$ groups to count tokens assigned to each group and then select corresponding $\frac{U[i]}{U[{i-1}]}$ groups to dispatch tokens. Similarly, each element $\mathcal{Z}_{a2a}^{\text{Intra-}(d\text{-}1)}[r,c,k]$ in $\mathcal{Z}_{a2a}^{\text{Intra-}(d\text{-}1)} \in \mathbb{R}^{E\times E\times G}$ denotes the number assigned to $k$-th expert group of size $G$ after swapping the positions of $r$-th and $c$-th experts before Intra-level-($d$-1) AlltoAll. 

Taking the configuration shown in Fig.~\ref{fig:hda} as the example where experts number and groups number are four and two, we use $\mathcal{Z}$ to show the duplicate-free number of tokens to two groups after swapping any two experts. After swapping E1 with E3, the duplicate-free number of tokens to two groups is both four, so $\mathcal{Z}[1,3,:]=\mathcal{Z}[3,1,:]=[4,4]$. Similarly, 
$\mathcal{Z}[2,4,:]=\mathcal{Z}[4,2,:]=[4,4]$, $\mathcal{Z}[{2,3,:}]=\mathcal{Z}[3,2,:]=[3,2]$ and $\mathcal{Z}[{1,4,:}]=\mathcal{Z}[{4,1,:}]=[2,3]$.

\begin{figure}[!t]
	\centering
 \begin{subfigure}[b]{0.24\textwidth}
		\centering
		\includegraphics[width=\textwidth]{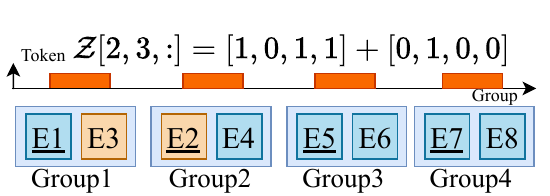}
		\caption{Case 1}
		\label{fig:case1}
	\end{subfigure}    
	 \begin{subfigure}[b]{0.24\textwidth}
		\centering
		\includegraphics[width=\textwidth]{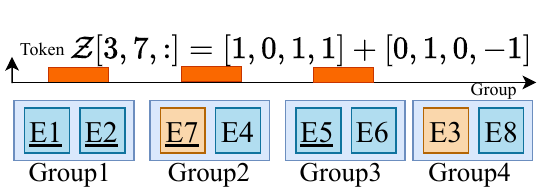}
		\caption{Case 2}
		\label{fig:case2}
	\end{subfigure}    
  \begin{subfigure}[b]{0.24\textwidth}
		\centering
		\includegraphics[width=\textwidth]{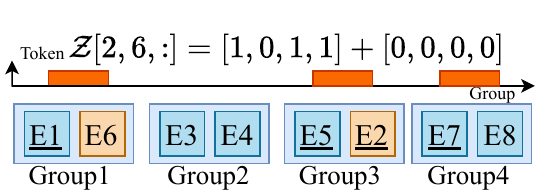}
		\caption{Case 3}
		\label{fig:case3}
	\end{subfigure}    
   \begin{subfigure}[b]{0.24\textwidth}
		\centering
		\includegraphics[width=\textwidth]{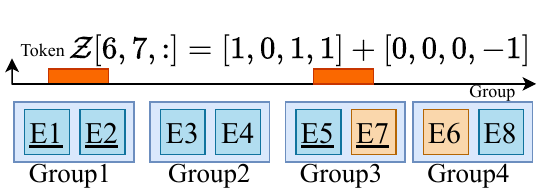}
		\caption{Case 4}
		\label{fig:case4}
	\end{subfigure}
	\caption{Four cases after swapping two experts with the one selected by the token while the other is not. An orange core indicates swapped experts, while an underline signifies experts selected by the token. }
	\label{fig:cases}
\end{figure}

However, directly calculating $\mathcal{Z}^{\text{Inter-}i}_{a2a}$ and $\mathcal{Z}_{a2a}^{\text{Intra-}(d\text{-}1)}$ is expensive, which requires a time complexity of $O(D\cdot T \cdot  K \cdot E^2)$ as the experts number can be large (256 in DeepSeek-V3\cite{liu2024deepseek} and 2048 in Switch\cite{DBLP:Fedus2022switch}), where $T$ is the total number of tokens for an MoE layer. 

To reduce the complexity, we design a strategy to calculate $\mathcal{Z}^{\text{Inter-}i}_{a2a}$ and $\mathcal{Z}_{a2a}^{\text{Intra-}(d\text{-}1)}$. 
Taking the configuration shown in Fig.~\ref{fig:cases} as the example where experts number and groups number are eight and four, we use $\mathcal{Z}$ to count the duplicate-free number of tokens to four groups after swapping any two experts. When a token arrives, it will select $K$ experts. If both swapped experts $A$ and $B$ are either selected or not by the token, the token number to each group remains unchanged, just as if there were no swapping. If one expert $A$ is selected while the other $B$ is not, there are four possible cases illustrated in Fig.~\ref{fig:cases}. In the first and second cases, if the group of the not selected expert $B$ has no selected experts (termed as ``Group2''), we must raise the ``Group2'' count after the swap. In the first case, if the group of the selected expert $A$ has at least two selected experts (illustrated as ``Group1''), no adjustment is needed for the ``Group1'' count. However, in the second case, with only one selected expert in the group (shown as ``Group4''), the ``Group4'' count must be decreased. In the third and fourth cases, if the group of the not selected expert $B$ has selected experts (termed as ``Group3''), the ``Group3'' count remains unchanged. For the group of the selected expert $A$, no change is required if there are at least two selected experts (illustrated as ``Group1''), as in the third case, but the value should be decreased if there is only one selected expert (shown as ``Group4''), as in the fourth case. 
Therefore, we initially assign $\mathcal{Z}^{\text{Inter-}i}_{a2a}$ and $\mathcal{Z}_{a2a}^{\text{Intra-}(d\text{-}1)}$ to the value without swapping and then adjust them across all cases to obtain the final value. The time complexity is reduced to $O(D\cdot T\cdot K\cdot E)$.




Then, we follow Eq.~(\ref{equ:optimal}) to get the final estimation matrix $\mathcal{Q}^{*}=\mathcal{Q}_{d^*}$ that represents the estimated time matrix of our HierD-AlltoAll after swapping the positions of any two experts.

\begin{theorem}
Given an MoE layer running on a cluster with expert parallelism using HierD-AlltoAll for communication, we can reduce the communication overhead by swapping the position of two experts. To achieve minimal communication time, the expert pair $(r^*,c^*)$ should satisfy
\begin{equation}
(r^*,c^*)=\operatorname*{arg\,min} \mathcal{Q}^{*}[{r,c}].
\label{theo:2}
\end{equation}
\begin{proof}
 As discussed in Eq.~(\ref{eq:estimate}) and Eq.~(\ref{equ:N}), we have covered all cases for swapping the position of two experts. Therefore, the optimal expert pair $(r^*,c^*)$ is identified by evaluating all cases to find the one minimizing communication time, i.e., $\operatorname*{arg\,min} \mathcal{Q}^*[{r,c}]$, which completes the proof.
\end{proof}
\end{theorem}


To improve the landscape of $\mathcal{Q}_d$, we choose a smoother max function\cite{zhou2015ell_1} to avoid abrupt changes in values as follows
\begin{equation}
  \text{smooth-max}(x, \gamma) = \max(x) \cdot \left( \sum_{i=1}^n \left( \frac{x[i]}{\max(x)} \right)^{\!\gamma} \right)^{\!1/\gamma},
  \label{equ:max}
\end{equation}
where $\gamma$ is a parameter to control the smoothness of the function.
We set $\gamma=10$ by default (will verify in \S\ref{subsec:max}).

\section{Evaluation}
\begin{table}[]
	\centering

 		\caption{The server configurations in our testbed.}
		\label{tab:server-config}

\begin{tabular}{ll}
\hline
\multicolumn{1}{l}{\textbf{Name}}    & \multicolumn{1}{l}{\textbf{Configuration}} \\ \hline
\multicolumn{1}{l}{CPU}     & \multicolumn{1}{l}{Dual Intel(R) Xeon(R) Platinum 8358 @ 2.60GHz} \\
\multicolumn{1}{l}{GPU}     & \multicolumn{1}{l}{8x Nvidia RTX A6000-48G @1.46GHz} \\
\multicolumn{1}{l}{Memory}  & \multicolumn{1}{l}{512GB DDR4} \\ 
\multicolumn{1}{l}{NVlink}  & \multicolumn{1}{l}{112.5GB/s (4x)} \\ 
\multicolumn{1}{l}{PCIe}    & \multicolumn{1}{l}{4.0 (x16)} \\ 
\multicolumn{1}{l}{Network} & \multicolumn{1}{l}{Mellanox MT28908 @ 200Gb/s} \\ \hline
\end{tabular}

\end{table}

\subsection{Experimental Settings}
\textbf{Testbeds.}
Experiments are carried out on a 32-GPU cluster comprising four interconnected nodes, each of which is equipped with eight Nvidia A6000 GPUs. The details of the server configuration are shown in Table~\ref{tab:server-config}. The software environments are Ubuntu-20.04, CUDA-12.1, PyTorch-2.1.2 and NCCL-2.18.5.

\textbf{Baselines. } We implement our HierMoE atop the prominent Megatron-LM training system, which supports various MoE models such as, DeepSeek and Qwen. We compare our HierMoE with three representative baselines Megatron-LM, SmartMoE and Tutel with 2DH-AlltoAll (Tutel-2DH).

\textbf{Real-World MoE Models.} To assess the end-to-end training performance on real-world MoE models, we exploit two commonly used MoE models based on DeepSeek-V3 and Qwen3-30B-A3B. Due to the GPU memory constraints of our testbed, we configure the hidden dimension and model dimension to be half of the original DeepSeek-V3 with 6 layers. For Qwen3-30B-A3B, we use 32 layers. For other configurations on the end-to-end experiments, we set micro batch size to 1, sequence length to 1024, the EP degree to 32, the same as the number of GPUs.

\begin{figure}[!t]
	\centering
	\begin{minipage}[b]{0.24\textwidth}
		\centering
		\includegraphics[width=\textwidth]{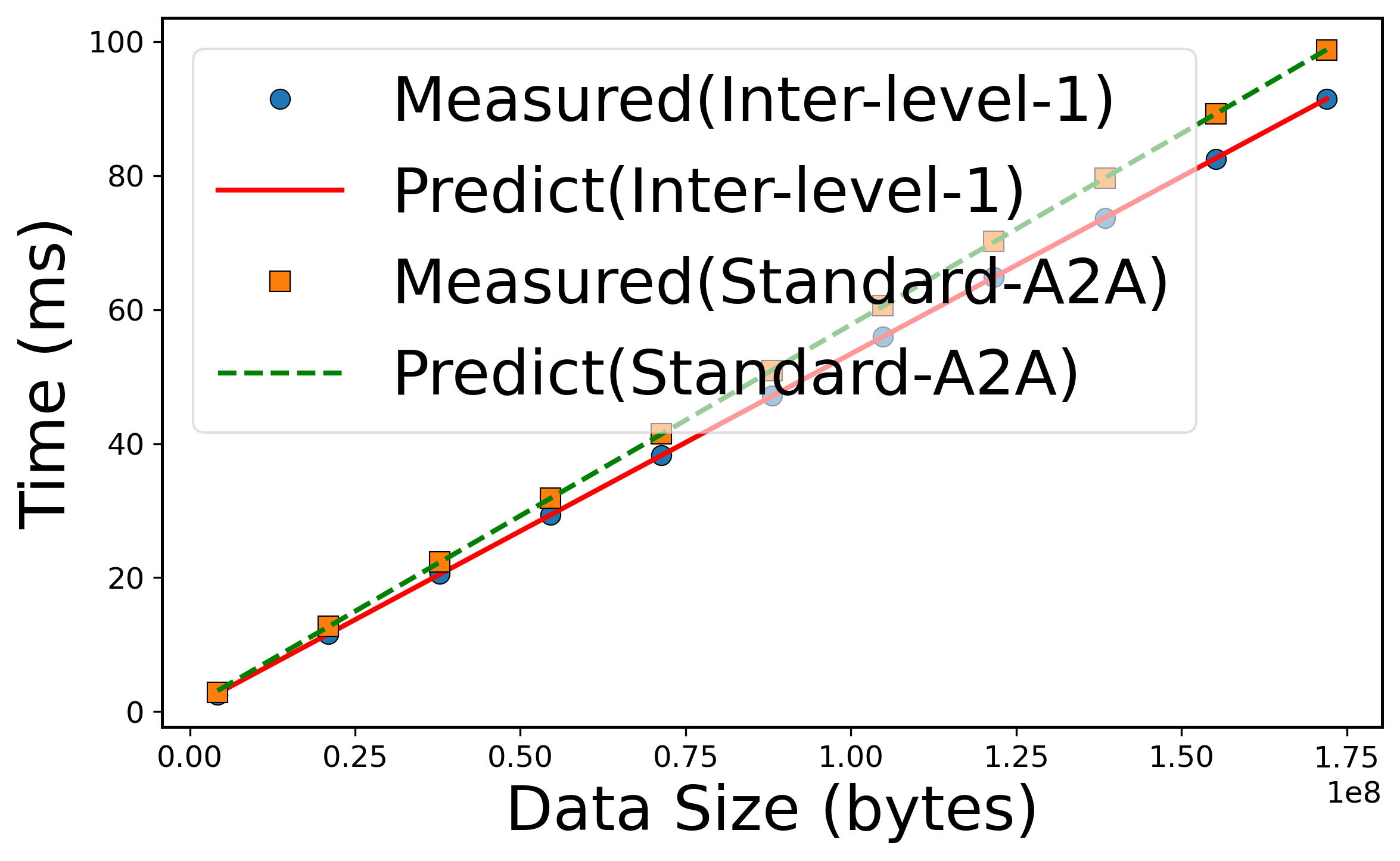}
		\caption*{(a) Inter-level-1 and standard AlltoAll.}
		
	\end{minipage}
	\hfill
	\begin{minipage}[b]{0.24\textwidth}
		\centering
		\includegraphics[width=\textwidth]{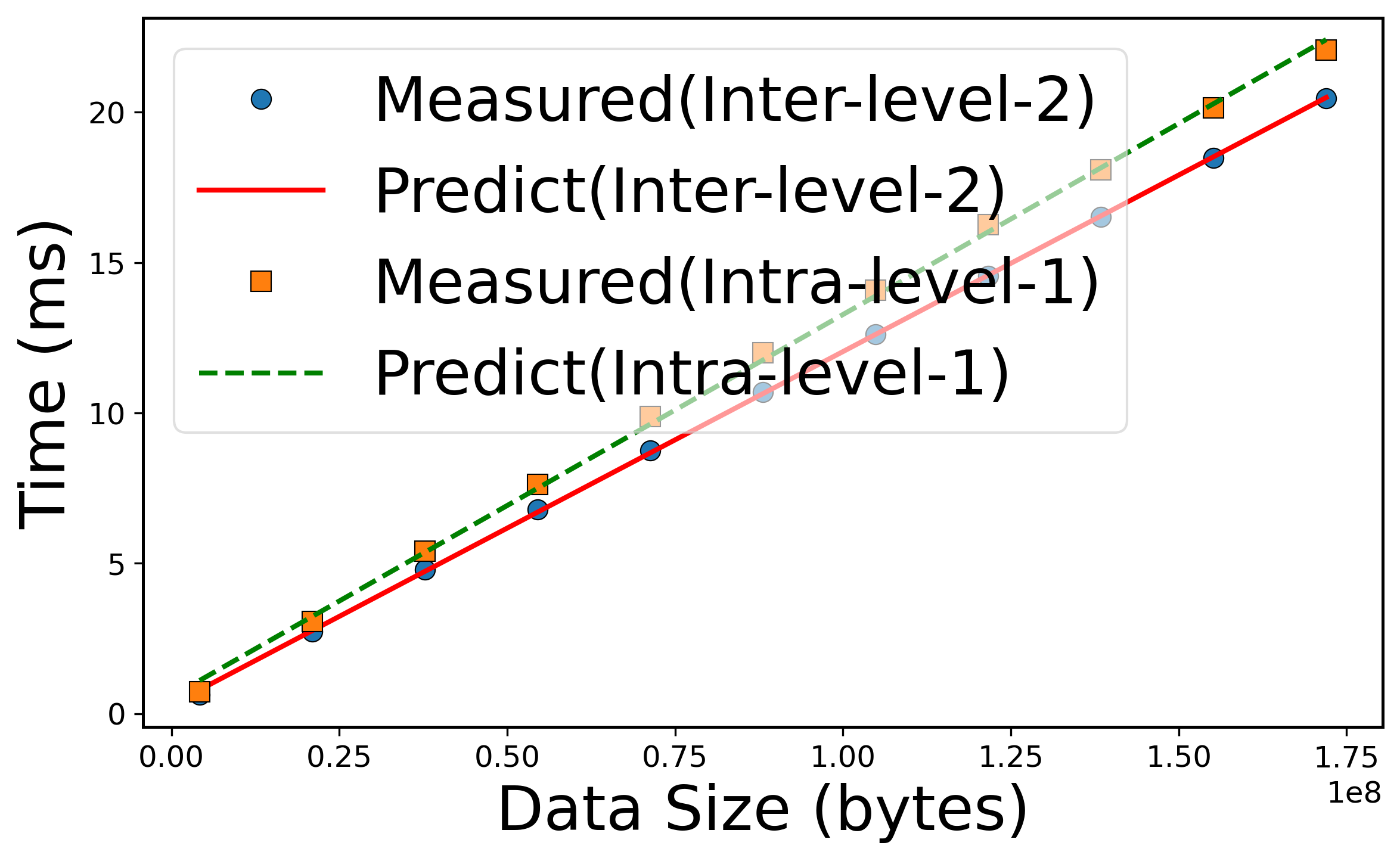}
		\caption*{(b) Intra-level-1 and Inter-level-2.}
	
	\end{minipage}
	
	\vspace{12pt}
	
	\begin{minipage}[b]{0.24\textwidth}
		\centering
		\includegraphics[width=\textwidth]{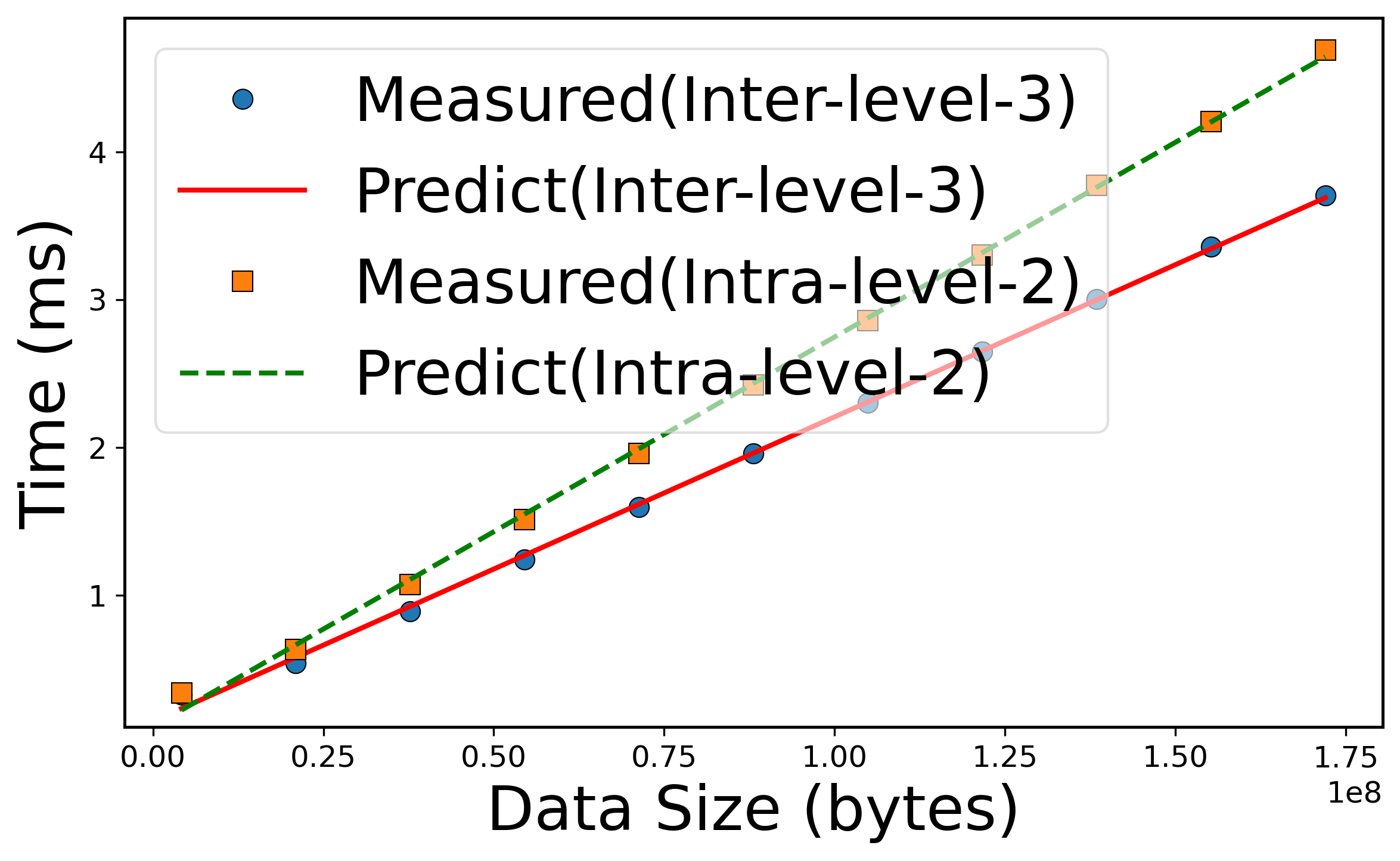}
		\caption*{(c) Intra-level-2 and Inter-level-3.}
	
	\end{minipage}
	\hfill
	\begin{minipage}[b]{0.24\textwidth}
		\centering
		\includegraphics[width=\textwidth]{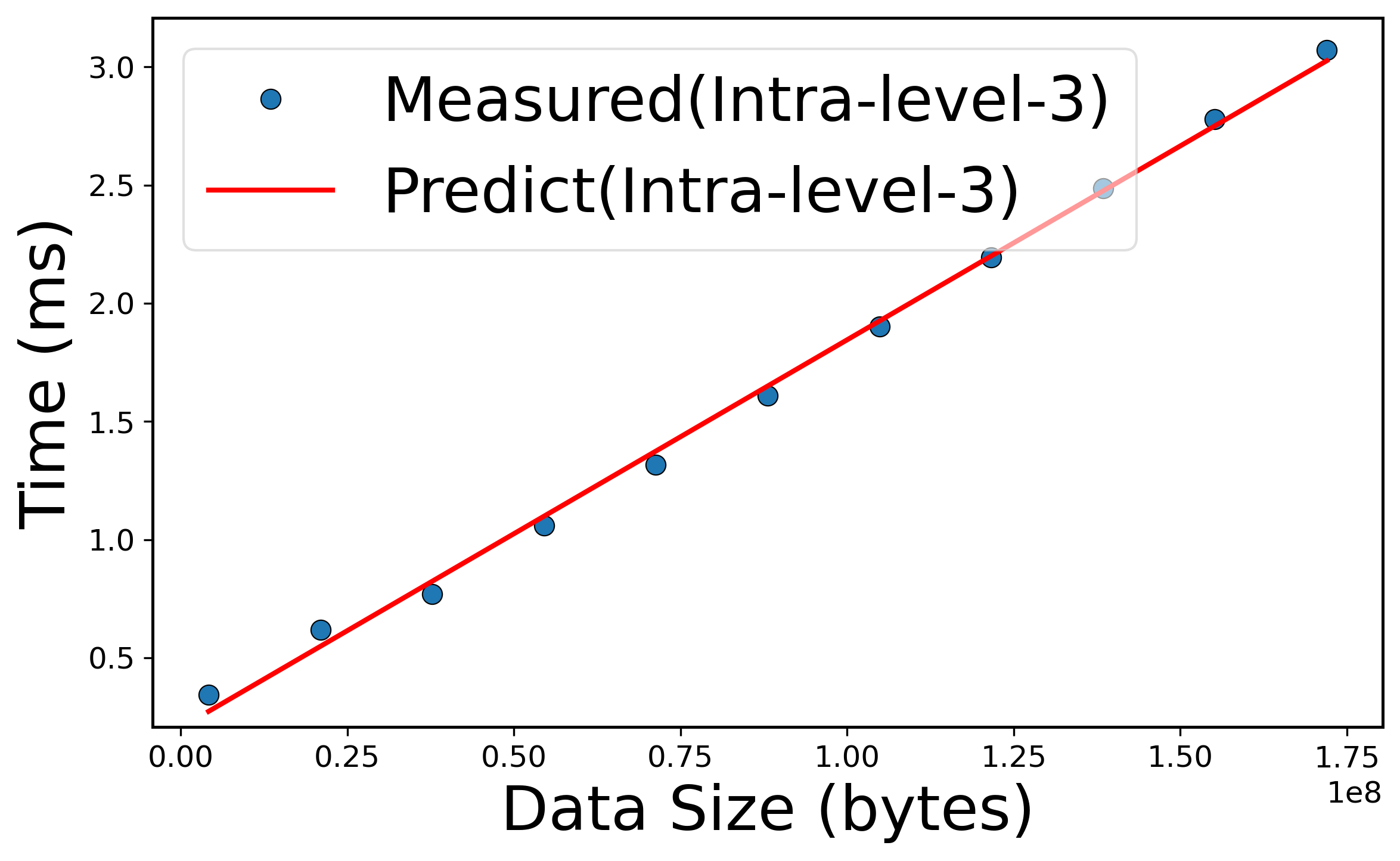}
		\caption*{(d) Intra-level-3 AlltoAll curve.}

	\end{minipage}
	
	\caption{Performance models. Markers are measured values and lines are predicted values with estimated parameters.
(a) $\alpha_{a2a}^{\text{inter(1)}} = 4.97 \times 10^{-1}$, $\beta_{a2a}^{\text{inter(1)}} = 5.29 \times 10^{-7}$,
$\alpha_{a2a} = 7.22 \times 10^{-1}$, $\beta_{a2a} = 5.70 \times 10^{-7}$.
(b) $\alpha_{a2a}^{\text{inter(2)}} = 3.01 \times 10^{-1}$, $\beta_{a2a}^{\text{inter(2)}} = 1.17 \times 10^{-7}$,
$\alpha_{a2a}^{\text{intra(1)}} = 5.71 \times 10^{-1}$, $\beta_{a2a}^{\text{intra(1)}} = 1.27 \times 10^{-7}$.
(c) $\alpha_{a2a}^{\text{inter(3)}} = 1.49 \times 10^{-1}$, $\beta_{a2a}^{\text{inter(3)}} = 2.06 \times 10^{-8}$,
$\alpha_{a2a}^{\text{intra(2)}} = 1.14 \times 10^{-1}$, $\beta_{a2a}^{\text{intra(2)}} = 2.63 \times 10^{-8}$.
(d) $\alpha_{a2a}^{\text{intra(3)}} = 2.04 \times 10^{-1}$, $\beta_{a2a}^{\text{intra(3)}} = 1.64 \times 10^{-8}$.}
	\label{fig:alphbeta}
\end{figure}

\subsection{Verification of Performance Models}\label{subsec:perfmodel-evaluation}
We require the input parameters that are related to the cluster for the performance models of AlltoAll communication. We measure the elapsed time with a range of sizes for seven types of AlltoAll communication to fit the performance models in Eq.~(\ref{pm}) and Eq.~(\ref{equ:level}) using micro-benchmark tools. In particular, we utilize the NCCL collective communication primitives along with \textit{nccl-tests}\footnote{\url{https://github.com/NVIDIA/nccl-tests}} to evaluate communication durations across diverse message sizes. As shown in Fig.~\ref{fig:alphbeta}, our linear models with intercept terms (i.e., startup time) can well fit the measured performance. Specifically, the \(r^2\) for the communication tasks are as follows: standard AlltoAll: 0.999997, Inter-level-1 AlltoAll: 0.999991, Intra-level-1 AlltoAll: 0.998922, Inter-level-2 AlltoAll: 0.998682, Intra-level-2 AlltoAll: 0.999051, Inter-level-3 AlltoAll: 0.999031, Intra-level-3 AlltoAll: 0.997245. The total time required for communication in the performance models is under 300 seconds. Fitting through the least squares method takes under 10 milliseconds.
When dealing with a new GPU cluster, it only needs to estimate the parameters one time using micro-benchmarks prior to model training, without impacting the training efficiency.

\begin{figure}[!t]
	\centering
		\includegraphics[width=0.35\textwidth]{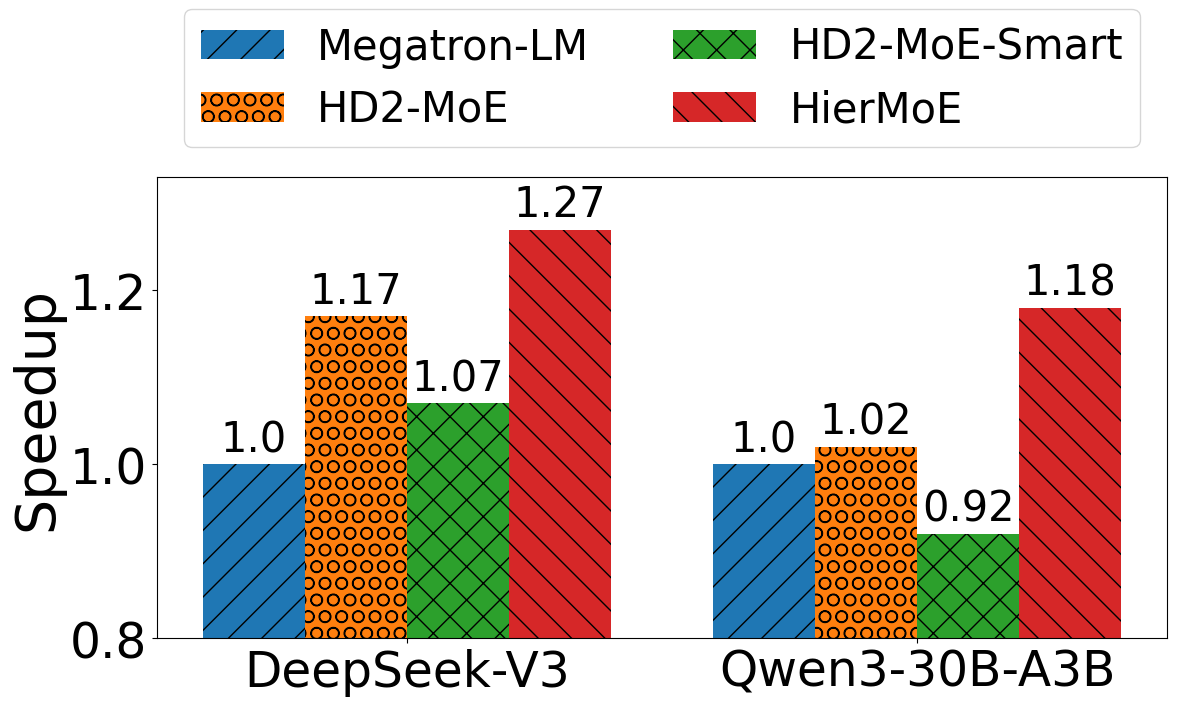}
	\caption{The end-to-end speedup ($\times$) of HierMoE, HD2-MoE and HD2-MoE-Smart over Megatron-LM on DeepSeek-V3 and Qwen3-30B-A3B.}
	\label{fig:rw}
\end{figure}

\begin{figure}[!t]
	\centering
		\includegraphics[width=0.35\textwidth]{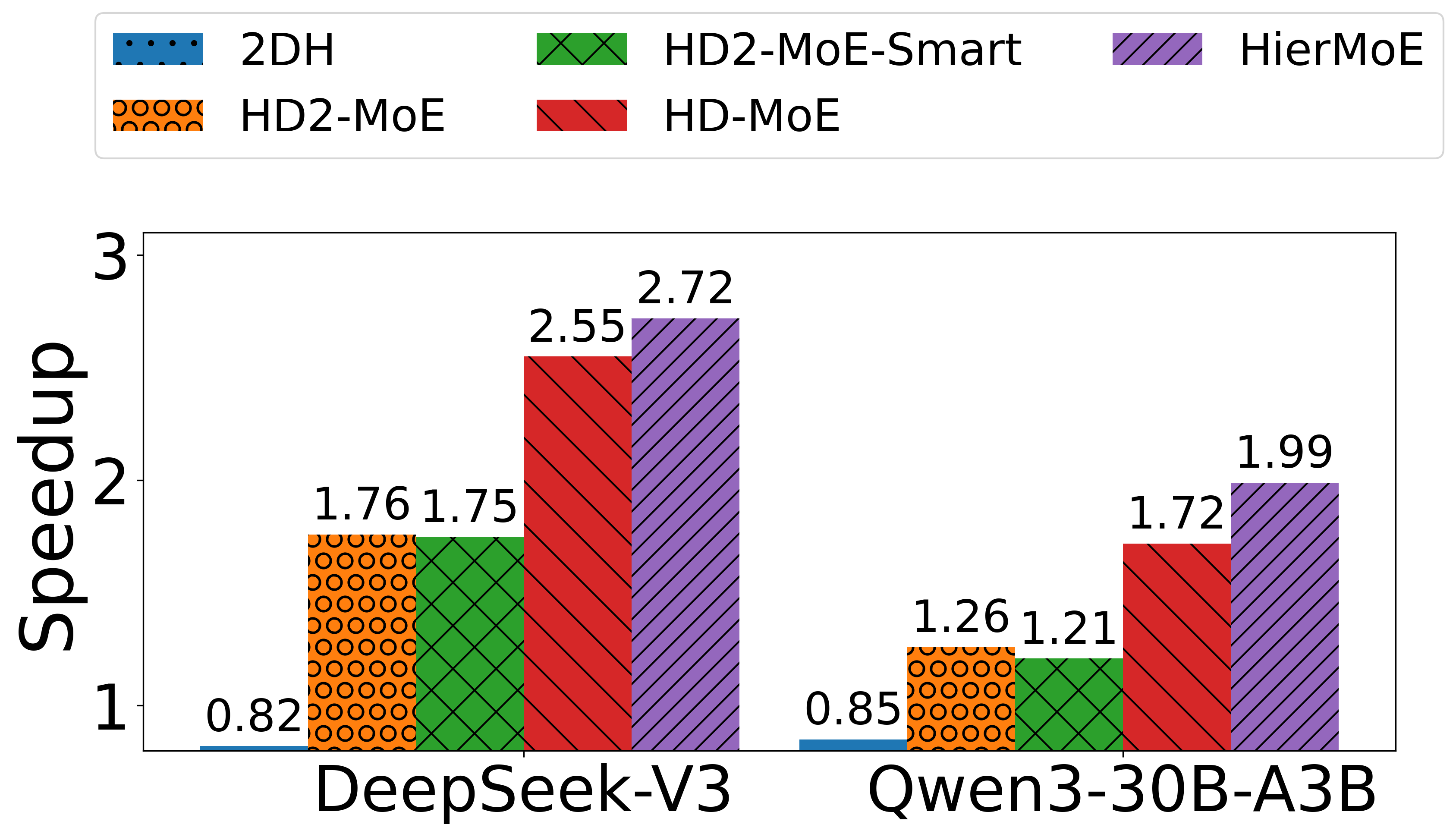}
	\caption{The AlltoAll communication speedup ($\times$) of Tutel-2DH, HD2-MoE, HD2-MoE-Smart, HD-MoE and HierMoE over Megatron-LM on DeepSeek-V3 and Qwen3-30B-A3B.}
	\label{fig:a2a}
\end{figure}
\subsection{End-to-end Training Time Comparison}
To evaluate the effectiveness of HierMoE, we compare HierMoE with Megatron-LM and SmartMoE on DeepSeek-V3 and Qwen3-30B-A3B models. For better comparison, we further perform experiments on an additional schedule, HD2-MoE, which only implements the two-dimensional hierarchical deduplication as shown in Fig.~\ref{fig:LEVEL1}. We also integrate our HD2-MoE with SmartMoE (termed as HD2-MoE-Smart). The experimental results are shown in Fig.~\ref{fig:rw}, which indicates that HierMoE achieves speedups of $1.18\times$ to $1.27\times$ compared to Megatron-LM. Additionally, HD2-MoE-Smart performs even worse than HD2-MoE, which validates that careful expert swap strategies are required on our HierD-AlltoAll. Compared to HD2-MoE, HierMoE can still achieve speedups of $1.10\times$ to $1.15\times$, validating the improvement of our HierD-AlltoAll and HierD-ES.

\subsection{AlltoAll Communication Time Comparison}
To further evaluate the effectiveness of HierMoE on AlltoAll communication. We compare the AlltoAll time of HierMoE with that of Megatron-LM, Tutel-2DH, HD2-MoE, HD2-MoE-Smart and HD-MoE (HierMoE w/o HierD-ES) on Deepseek-V3 and Qwen3-30B-A3B as shown in Fig.~\ref{fig:a2a}. The experimental results reveal that our HierMoE provides $1.55 \times$ to $1.64 \times$ speedups over HD2-MoE-Smart, $1.99 \times$ to $2.72 \times$ speedups over Megatron-LM, and $2.34\times$ to $3.32 \times$ over Tutel-2DH, showing the effectiveness of our approach. 
It can be seen that
Tutel-2DH performs worse than Megatron-LM, whereas our HD2-MoE achieves $1.26\times$ to $1.76\times$ speedups over Megatron-LM. Furthermore, HD2-MoE-Smart is less effective than HD2-MoE, illustrating the limitations of SmartMoE. Furthermore, HD-MoE achieves a speedup of $1.37\times$ to $1.45 \times$ compared to HD2-MoE, demonstrating the efficacy of our HierD-AlltoAll. In addition, HierMoE boosts the performance by $2.55\times$ to $2.72\times$ on DeepSeek-V3 and $1.72\times$ to $1.99\times$ on Qwen3-30B-A3B by implementing HierD-ES atop HierD-AlltoAll, further highlighting the importance of HierD-ES. 

\begin{figure}[!t]
	\centering
		\includegraphics[width=0.38\textwidth]{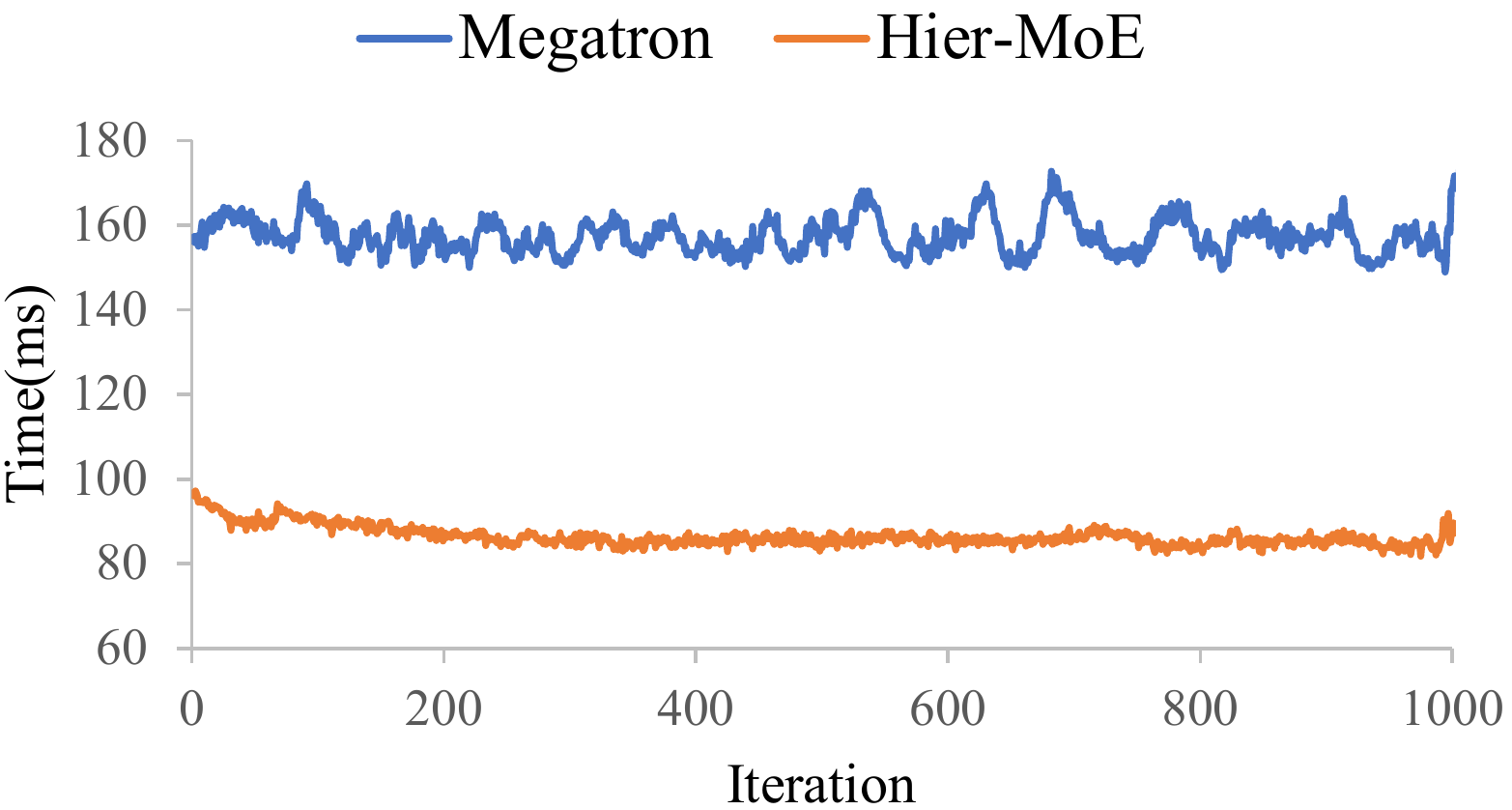}
	\caption {The curve's smoothness comparing the time cost of AlltoAll for HierMoE and Megatron as iterations rise at the first layer of Qwen3-30B-A3B.}
	\label{fig:line}
\end{figure}

Furthermore, we assess the iteration time during the training iterations as shown in Fig.~\ref{fig:line}. It is seen that our HierMoE is much more stable than that of Megatron-LM.

\subsection{Ablation Study}

\begin{figure}[!t]
	\centering
		\includegraphics[width=0.4\textwidth]{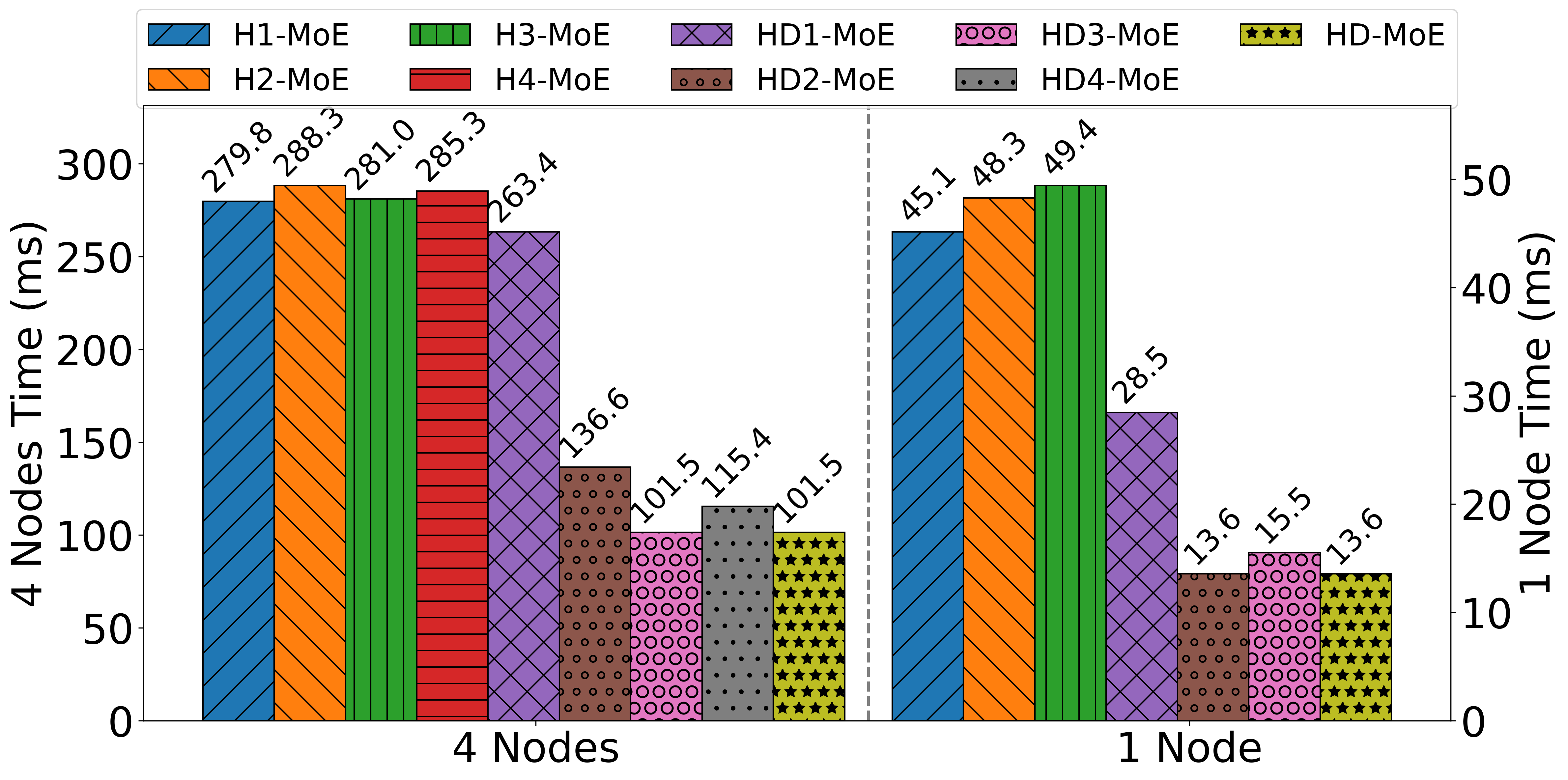}
	\caption {The time cost of AlltoAll for different configurations on 4 nodes and 1 node.}
	\label{fig:hvhd}
\end{figure}

\begin{table}[]
	\centering

 		\caption{The AlltoAll communication speedup ($\times$) of HD2-MoE, HD-MoE and HierMoE over Megatron-LM with varied $K$, $E$, and $G$.}
		\label{tab:vary}

\setlength{\tabcolsep}{4pt}
\begin{tabular}{@{} l *{3}{S[table-format=1.2]} *{3}{S[table-format=1.2]} *{3}{S[table-format=1.2]} @{}}
\toprule
Method & \multicolumn{3}{c}{$K$} & \multicolumn{3}{c}{$E$} & \multicolumn{3}{c}{$G$} \\
\cmidrule(lr){2-4} \cmidrule(lr){5-7} \cmidrule(lr){8-10}
 & {6} & {\bfseries 8} & {10} & {64} & {\bfseries 128} & {256} & {8} & {16} & {\bfseries 32} \\
\midrule
HD2-MoE & 0.95 & 1.26 & 1.30 & 1.12 & 1.26 & 1.18 & 1.24 & 1.71 & 1.26 \\
HD-MoE  & 1.37 & 1.72 & 1.82 & 1.57 & 1.72 & 1.61 & 2.36 & 2.50 & 1.72 \\
HierMoE & 1.56 & 1.99 & 2.10 & 1.83 & 1.99 & 1.84 & 2.86 & 2.62 & 1.99 \\
\bottomrule
\end{tabular}
\end{table}

\textbf{Impacts by $K$, $E$, and $G$. }
We evaluate the performance with configured different $K$ (the number of top experts selected for each token), $E$ (the number of experts per MoE layer), and $G$ (the total number of GPUs of the cluster) and measure the speedups of our proposed methods over Megatron-LM as shown in Table~\ref{tab:vary}. Results indicate that HierMoE achieves speedups ranging from $1.83\times$ to $1.99\times$ with different $E$. And the speedup of  HierMoE improves from $1.56\times$ to $2.10\times$ as $K$ increases, highlighting a worsened token duplication issue with higher $K$. As $G$ rises, the speedup of HierMoE drops. At $G=8$, without inter-node communication, its speedup distinctly differs from the others. For $G\in\{16,32\}$, the increase in nodes leads to reduced duplicate rates at the first hierarchical level, resulting in a decrease in HD2-MoE's speedup. Nevertheless, HD-MoE achieves a comparable speedup with HD2-MoE with $1.38\times$ when $G=16$ and $1.37\times$ when $G=32$, demonstrating the validity of our approach in determining the dimensions for HierD-AlltoAll. Meanwhile, compared to HD2-MoE, the speedup of HierMoE improves from $1.53\times$ to $1.58\times$ with increasing $G$ from $16$ to $32$, highlighting the effectiveness of our HierD-ES strategy.

\textbf{Performance with different dimensions.}
We assess the AlltoAll time cost for nine configurations to determine the influence of varying dimensions on 4 nodes and 1 node, as depicted in Fig.~\ref{fig:hvhd}. H1-MoE, H2-MoE, H3-MoE, and H4-MoE denote the MoE layer using hierarchical AlltoAll without deduplication, whereas HD1-MoE, HD2-MoE, HD3-MoE, and HD4-MoE represent the MoE layer with hierarchical deduplication AlltoAll. Notably, experiments on 1 node only have 3 dimensions. HD-MoE corresponds to the MoE layer employing our HierD-AlltoAll. Results indicate that while hierarchical AlltoAll does not reduce communication overhead, our deduplication approach does, and our HierD-AlltoAll optimally selects the dimension.

\textbf{Performance with different kinds of max functions. }\label{subsec:max}
We also conduct a set of experiments over three kinds of max function on Eq. (\ref{equ:N}), including a smooth max function on Eq. (\ref{equ:max}), a standard max function and a Log-Sum-Exp function ($\ln{\sum_i(\exp{x[i]}})$). Results show that the standard max function, smooth max function and Log-Sum-Exp achieve speed up HierMoE over HD-MoE by $1.13\times$, $1.17 \times$ and $1.16 \times$.
Smooth max functions improve little to the performance. We thus simply choose the best one. 
We also evaluate HierMoE against HD-MoE by varying $\gamma$ within $[5,7,9,11,13,15,17,19]$ to assess sensitivity to the max function's smoothness in Eq. (\ref{equ:max}). Results indicate a speed up of HierMoE over HD-MoE between $1.16\times$ and $1.17\times$, suggesting low sensitivity to $\gamma$.

\textbf{Performance with varied expert placements updating frequency. }
In practice, swapping two experts takes just 1\% of the total end-to-end time. We find that HierMoE achieves $1.17\times$, $1.17\times$, $1.15\times$, and $1.13\times$ faster than HD-MoE with an HierD-ES update frequency of every 1, 2, 4, and 8 iterations, respectively. Higher frequencies are seen to have better performance, so we choose to update HierD-ES every iteration.
\section{Conclusion}
In this work, we present HierMoE, a novel MoE training approach that substantially reduces communication overhead through three key innovations: 1) a hierarchical All-to-All mechanism with token deduplication that eliminates redundant transfers across hierarchical levels, 2) a hierarchical expert placement strategy to align diverse efficiencies across different hierarchical levels in our proposed All-to-All, and 3) theoretical models to evaluate and enhance the hierarchical deduplication All-to-All and expert replacement strategy. Implemented atop Megatron-LM, our HierMoE demonstrates significant performance gains with extensive experiments conducted on a 32-GPU cluster using DeepSeek-V3 and Qwen3-30B-A3B models, achieving  $1.55\times$ to $3.32\times$ faster AlltoAll communication compared to state-of-the-art systems like Tutel-2DH, SmartMoE and Megatron-LM, while delivering $1.18\times$ to $1.27\times$ faster end-to-end training time. 

\section*{Acknowledgments}
The research was supported in part by the National Natural Science Foundation of China (NSFC) under Grant No. 62302123, Grant No. 62272122, and Grant No. 62376073, Guangdong Provincial Key Laboratory of Novel Security Intelligence Technologies under Grant 2022B1212010005, the Guangzhou Municipal Joint Funding Project with Universities and Enterprises under Grant No. 2024A03J0616, Guangzhou Municipality Big Data Intelligence Key Lab (2023A03J0012), Shenzhen Science and Technology Program under Grant No. KJZD20240903104103005, Grant No. KJZD20230923114213027 and Grant No. KJZD20230923115113026, the Colleges and Universities Stable Support Project of Shenzhen, China (No. GXWD20220817164856008 and No. GXW-
D20220811173149002), Hong Kong RGC CRF grants under contracts C7004-22G and C6015-23G.
\bibliography{ref}
\bibliographystyle{IEEEtran}

\end{document}